\documentclass{aa}
\usepackage{natbib}
\bibpunct{(}{)}{;}{a}{}{,} 
\input{bib.def}
\usepackage{hyperref}
\hypersetup{colorlinks=true,urlcolor=blue,citecolor=blue,linkcolor=blue}
\usepackage{txfonts}

\usepackage{xcolor,colortbl}

\usepackage{graphicx}
\usepackage{txfonts}
\usepackage{caption}
\usepackage{subcaption}
\usepackage{ulem}
\usepackage{booktabs}
\usepackage{multirow}
\begin{document} 

   \title{The Sun at millimeter wavelengths}
   \subtitle{IV. Magnetohydrodynamic waves in small-scale bright features}   

   \author{Juan Camilo Guevara G\'omez \inst{1,2}
          \and
           Shahin Jafarzadeh  \inst{3,1} 
          \and
          Sven Wedemeyer  \inst{1,2}
          \and
          Samuel D. T. Grant \inst{4}
          \and
          Henrik Eklund \inst{5,1,2}
          \and
          Mikołaj Szydlarski \inst{1,2}}
   
  \authorrunning{Guevara G\'omez {et~al.}}
  \titlerunning{The Sun at millimeter wavelengths. IV.}
   \institute{Rosseland Centre for Solar  Physics, University of Oslo, Postboks 1029 Blindern, 0315 Oslo, Norway\\
   \email{j.c.g.gomez@astro.uio.no}
            \and
            Institute of  Theoretical Astrophysics, University of Oslo, Postboks 1029 Blindern, 0315 Oslo, Norway 
            \and
            Max Planck Institute for Solar System Research, Justus-von-Liebig-Weg 3, 37077 G\"{o}ttingen, Germany
            \and
            Astrophysics Research Centre, School of Mathematics and Physics, Queen’s University Belfast, Belfast, BT7 1NN, UK
            \and
            Institute for Solar Physics, Department of Astronomy, Stockholm University AlbaNova University Centre, SE-106 91 Stockholm, Sweden\\
} 
 
\def\corrAuthor{JCGG}

   \date{Received 10 June 2022 ; accepted 20 December 2022 }

\abstract{}
{We used solar observations of a plage/enhanced network with the Atacama Large Millimeter/sub-millimeter Array (ALMA) in Band 3 and Band 6 together with synthetic continuum maps from numerical simulations with Bifrost at the same bands to carry out a detailed study of bright small-scale magnetic features.}
{We have made use of an algorithm to automatically identify and trace bright features within the field of view (FoV) of the ALMA observations and the simulation. In particular, the algorithm recovers information of the time evolution of the shape, motion of the centre of gravity, temperature and size for each feature. These quantities are used to determine the oscillatory properties of each feature utilising wavelets analysis.}
{We found {193 and 293} features in the Bands 3 and 6 observations, respectively. In the {degraded} simulation, the total number of features were {24} for Band 3 and {204} for Band 6. {In the original simulation, the total number of features were 36 for Band 3 and 392 for Band 6.} Based on the simulation, we confirm the magnetic nature of the features. We have obtained average oscillation periods of {30-99}\,s for temperature, {37-92}\,s for size and {37-78}\,s for horizontal velocity. There are indications for the  possible presence of transverse (kink) waves with average amplitude velocities of {2.1-5.0}\,km\,s$^{-1}$. We find a predominant anti-phase behaviour between temperature and size oscillations suggesting that the variations of the bright features are caused by compressible fast-sausage Magnetohydrodynamics (MHD) modes. We, for the first time to our knowledge, estimate the flux of energy of the fast-sausage waves at the chromospheric heights sampled by ALMA as {453-1838\,W\,m$^{-2}$} for Band 3 and {3640-5485\,W\,m$^{-2}$} for Band 6.}
{We have identified MHD waves, both transverse (kink) and compressible sausage modes, in small-scale (magnetic) structures, independently, in both ALMA Band 3 and Band 6 observations, and their corresponding synthetic images from simulations. The decrease of wave energy-flux with height (from Band 6 to Band 3) could possibly suggest energy dissipation at chromospheric heights, thus wave heating, with the assumptions that the identified small-scale waves are typical at each band and they propagate upward through the chromosphere.
}


   \maketitle

\section{Introduction}
\label{sec:introduction}

The solar atmosphere is replete with distinct types of waves caused by perturbations of the plasma \citep{1974ARA&A..12..407S}. In particular, the magnetic field, ubiquitously present as diverse structures in the solar atmosphere, acts as an effective restoring force for perturbations, giving rise to different types of magnetohydrodynamic (MHD) waves \citep[see the reviews by][]{2005SSRv..121..115N,2009SSRv..149..299V,2009SSRv..149...65D,2015SSRv..190..103J, 2016GMS...216..449J, 2022LRSP...00..000J}. The magnetic field also provides an optimal wave conduit, as compressible and incompressible MHD-wave modes are readily generated in magnetic flux structures by recurring perturbations around the solar surface \citep{2016GMS...216..431V}. These MHD waves have long been suggested as playing a key role in the transport of energy and the resulting  heating of the upper solar atmosphere \citep[e.g.,][]{1961ApJ...134..347O, 1994ApJ...435..482P, 2006SoPh..234...41K, 2021JGRA..12629097S}. The focus of current research lies in assessing the energy budget of waves, and expanding on our tentative candidates for dissipation of wave energy into local plasma heating \citep[i.e.,][]{1978ApJ...226..650I, 2011ApJ...736....3V, 2018NatPh..14..480G}.

Compressible MHD wave modes are determined by their ability to perturb the local plasma density. An example of these are the compressible MHD sausage mode, whose observational signatures are periodic fluctuations in the cross-sectional area of their cylindrical waveguide, accompanied by periodic fluctuations in intensity. The mode of the oscillation can be discerned from the phase relationship between the cross-sectional area and total intensity oscillations, with slow sausage modes displaying an in-phase behaviour, and an anti-phase relationship for fast sausage modes \citep{1983SoPh...88..179E,2013A&A...555A..75M}. With the advent of high-resolution instruments such as the Rapid Oscillations in the Solar Atmosphere imager \citep[ROSA;][]{2010SoPh..261..363J}, in-depth studies of sausage modes in the lower solar atmosphere could be conducted \citep[e.g.,][]{2011ApJ...729L..18M}. Sausage modes have since been positively identified in a variety of magnetic structures, such as sunspots \citep{2014A&A...563A..12D}, magnetic pores \citep{2015ApJ...806..132G, 2018ApJ...857...28K, 2021RSPTA.37900172G} and chromospheric fibrils \citep{2012NatCo...3.1315M, 2017ApJS..229....7G}. Recently, \cite{2021RSPTA.37900184G} utilised the unique observations of the Atacama Large Millimeter/submillimeter Array \citep[ALMA;][]{2009IEEEP..97.1463W} to reveal the signatures of fast sausage modes in small-scale features previously undetectable by intensity and velocity observations.

Incompressible MHD (transverse) kink modes manifest as motions perpendicular to the vertical waveguide axis. If the cross-sectional area of the waveguide is perpendicular to the line of sight, these modes are observed as horizontal motions with an oscillatory pattern. Various structures in the photosphere and chromosphere such as small-scale bright points, fibrils, mottles and spicules have been identified as waveguides in which kink modes are excited, having periods on the order of 30-350~s and velocity amplitudes of about 1-29~km\,s$^{-1}$  \citep[e.g.,][]{2007Sci...318.1574D,2007SoPh..246...65L,2012ApJ...750...51K,2013A&A...554A.115S, 2014A&A...569A.102S,2015SSRv..190..103J,2015A&A...577A..17S,2017ApJ...840...19S,2017ApJS..229....9J,2017ApJS..229...10J}). \citet{2021RSPTA.37900184G} also observed transverse oscillations in three small-scale bright features observed in solar ALMA observations with periods on the order of 30-136~s and velocity amplitudes between 0.6~km\,s$^{-1}$ and 5.9~km\,s$^{-1}$, although the structures associated with the waveguides were not identified.  

Evaluating atmospheric plasma parameters, such as temperature, is essential to calculate the energy flux of MHD waves. The inherent complexities in spectral line formation in the chromosphere result in simplified estimations of these parameters, such as interpolated models \citep{2015ApJ...806..132G}, or photospheric inversions \citep{2021RSPTA.37900172G}. However, the introduction of solar ALMA observations provides a novel insight into the chromosphere, through observations of millimetre continuum radiation formed mostly by thermal bremsstrahlung (free-free emission) \citep[see, e.g.,][and references therein]{2021MNRAS.500.1964V,2016SSRv..200....1W,2006A&A...456..697W,1985ARA&A..23..169D}. ALMA records brightness temperature, which is analogous to the actual gas temperature of the continuum forming layer, providing unprecedented accuracy in chromospheric plasma temperature measurements. ALMA observations can also be tailored for observing oscillations and waves in the solar atmosphere, with high temporal cadences of up to 1\,s 
\citep[see, e.g.,][]{2021RSPTA.37900174J,2020A&A...634A..86P,Narang2020,2021A&A...652A..92N, 2020A&A...644A.152E}.

For this study, we utilise high cadence observations of brightness temperatures with ALMA for the detection and analysis of wave modes in the solar atmosphere, supported by artificial millimetre observations based on numerical simulations. The ALMA data and the numerical simulations are described in Sect.~\ref{sec:data}, followed by a description of detection method for oscillation events in Sect.~\ref{sec:methods}. The analysis of the detected events and exhibited wave modes are presented in Sect~\ref{sec:analysis}. A discussion and conclusions are provided in Sect.~\ref{sec:discussion}.


\begin{figure*}[tp!]
    \centering
    \includegraphics[width=.95\textwidth]{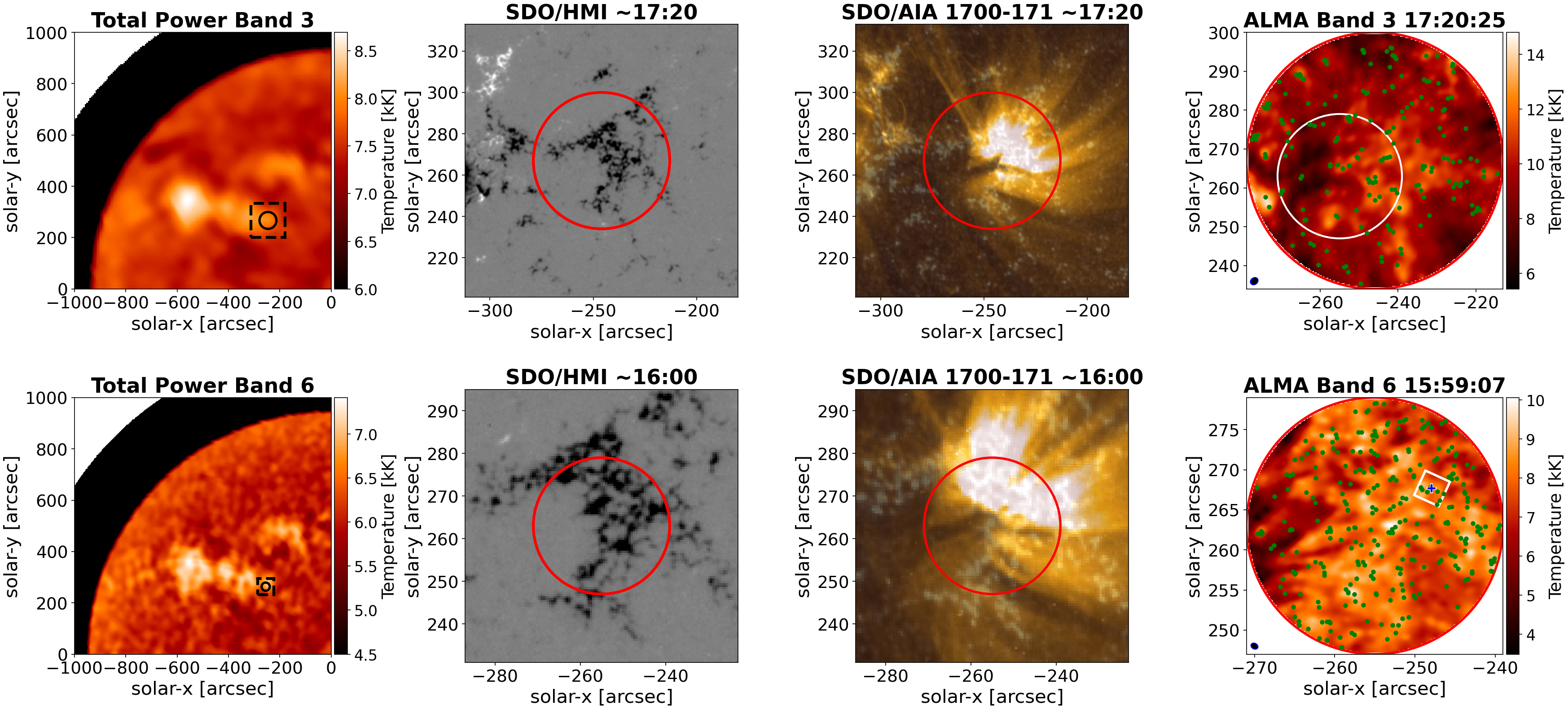}
    \caption{Top row from left to right: total power {or full-disk} map in ALMA Band 3 in which the small black dashed square delimits the region showed in SDO/HMI and SDO/AIA 1700-171 and the small black circle depicts the ALMA Band 3 interferometric FoV, SDO/HMI image showing the presence of magnetic field in the ALMA FoV indicated by the red circle, SDO/AIA 170- 17.1\,nm composite image of the same region, ALMA Band 3 image in which the green dots indicate the median location of all analysed feature and the white circle shows the ALMA Band 6 FoV. Bottom row: same as in top row but for Band 6, in the ALMA Band 6 image the small white square with the blue cross in the centre signalise the location of the example feature used in this paper. The small black ellipse in the bottom-left corner of ALMA images show their respective beam shapes.\label{fig:context_sdo}}
\end{figure*}

\begin{figure}[ht]
    \centering
    \includegraphics[width=.49\textwidth]{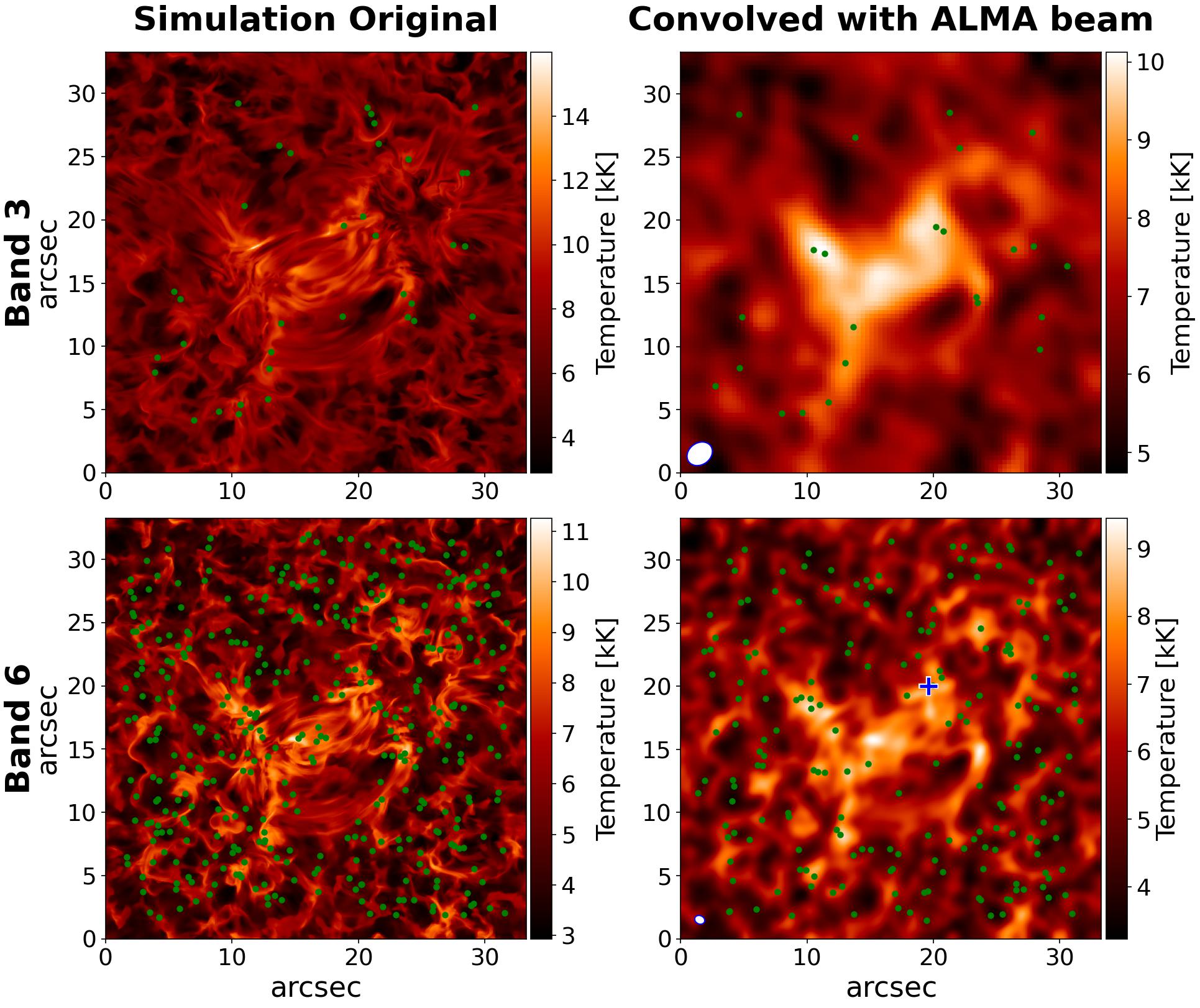}
    \caption{Same snapshot from Bifrost simulation for the synthetic ALMA Band 3 and 6 in the top and bottom row respectively. Left column corresponds to original snapshots with full resolution. Right column shows the result of convolving and matching with ALMA Beam and pixel size. The green dots mark the median position of all analysed features. The blue cross by (20,20) arcsec {in the Band 6 degraded simulation} shows the position of the example feature.\label{fig:sim_convbeam}}
\end{figure}

\section{Data}
\label{sec:data}

In this study, solar ALMA observations are employed alongside synthetic continuum maps from a Bifrost simulation{, both in the original spatial resolution and in a} degraded {resolution} to match ALMA's spatial and temporal resolution.

\subsection{Observations}
\label{sec:observations}

\begin{figure*}[tp!]
    \centering
    \includegraphics[width=0.95\textwidth]{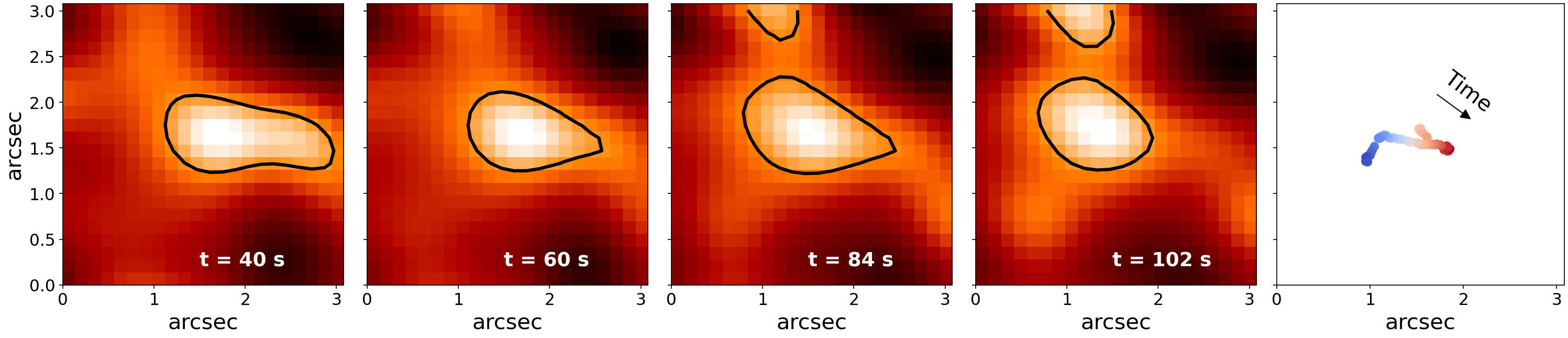}
    \caption{Time evolution of a feature in Band 6 observation. The 4 panels correspond to times marked as vertical lines in the top-left panel of Figure~\ref{fig:featexam_cwt}. In each panel the black contours depict the feature border found by thee algorithm. The most right panel shows the trajectory of the centre of gravity with the arrow indication the direction of time.\label{fig:featexamp_snaps}}
\end{figure*}

\begin{figure}[ht]
    \sidecaption
    \includegraphics[width=.45\textwidth]{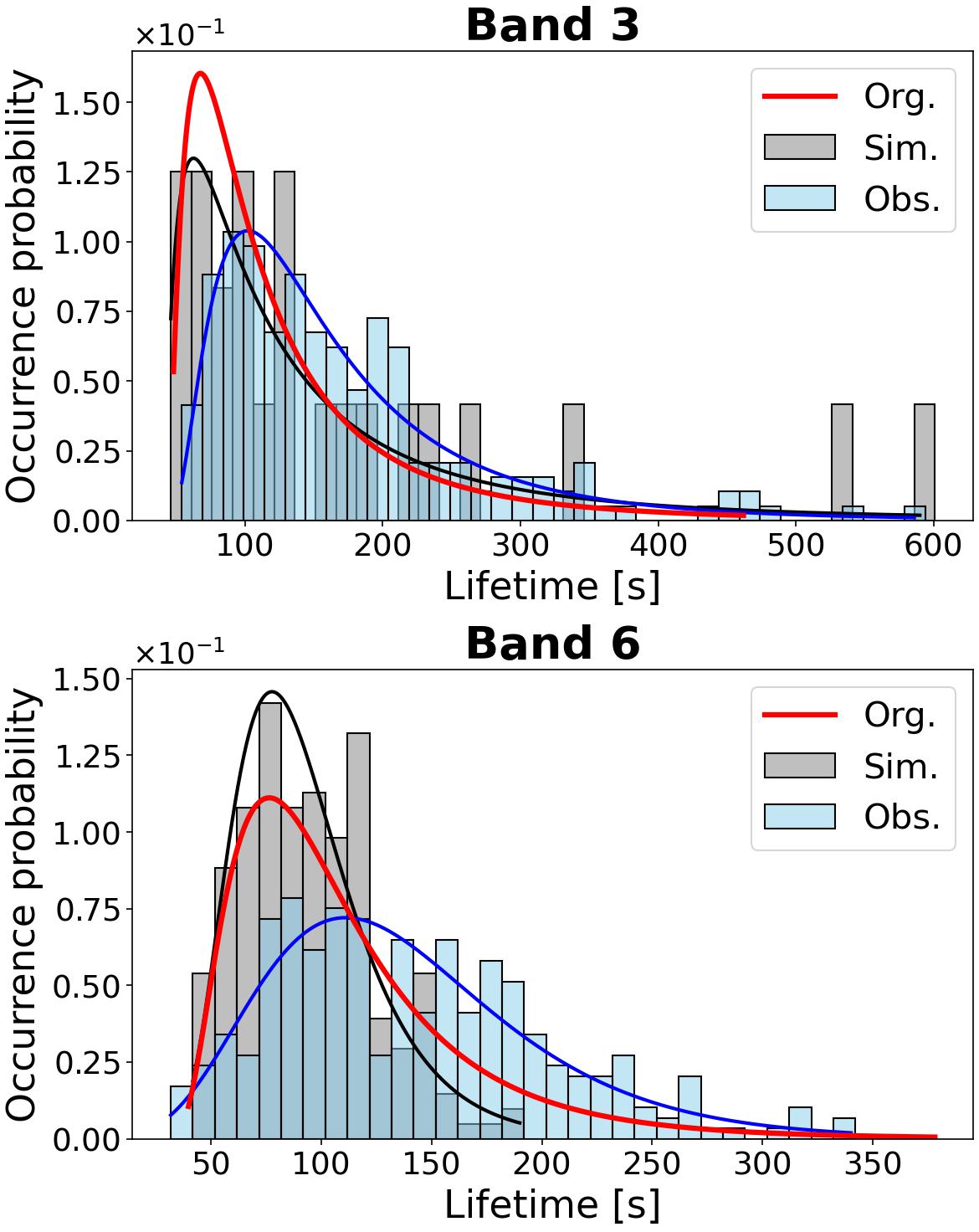}
    \caption{{Distribution of lifetimes for all the features for Band 3  (top) and Band~6 (bottom). In both panels the blue histograms correspond to results from observations and the grey to results from degraded simulations. The solid lines are their corresponding log-norm fits. The red solid line corresponds to the log-norm fit of the results from the simulation at the original resolution.\label{fig:Lifes}}}
\end{figure}

The ALMA Band 3 (2.8-3.3 mm) and Band 6 (1.2-1.3 mm) observations were carried out on 22nd April 2017, as detailed in \citet{2021RSPTA.37900174J}. Band 6 observations took place from 15:59 to 16:44 UTC followed by Band 3 observation between 17:20 and 17:55 UTC both as part of the program 2016.1.00050.S. The reported solar P-angles, which determine the rotation of the ALMA field of view (FoV) with respect to the solar north, were $-25.32^\circ$ and $-24.65^\circ$ for bands 3 and 6, respectively. Each of the observed bands consist of four sub-bands which are centred at different wavelengths within the band range \citep[see, e.g.,][]{2020A&A...635A..71W,2020A&A...644A.152E}. For the purpose of this study, we have used the full-band image reconstruction mode in which all the sub-bands of a particular band are combined into one, resulting in time series for frequencies around 3 mm for Band 3 (100 GHz) and 1.25 mm (240 GHz) for Band 6. The full-band ALMA images have a higher signal-to-noise ratio (S/N) compared to the sub-bands S/N because of the better sampling of the spatial Fourier space during the image reconstruction. As an interferometer, ALMA measures the temperature differences on the observed source. It is therefore necessary to use the additional Total Power (TP) maps to obtain the absolute temperature values for the observed source. {The TP maps are full-disk images of the Sun that are derived from observations with the (single-dish) total power antennas of ALMA using the fast scanning technique described by \citet{2017SoPh..292...88W}}. The absolute temperature ranges of the ALMA observations are [5106,15126]\,K with a mean of 9317\,K and a standard deviation of 1229\,K for Band~3 and [2090,12196]\,K with a mean of 7496\,K and a standard deviation of 1014\,K for Band~6. 

\begin{figure}[h!]
    \sidecaption
    \includegraphics[width=.45\textwidth]{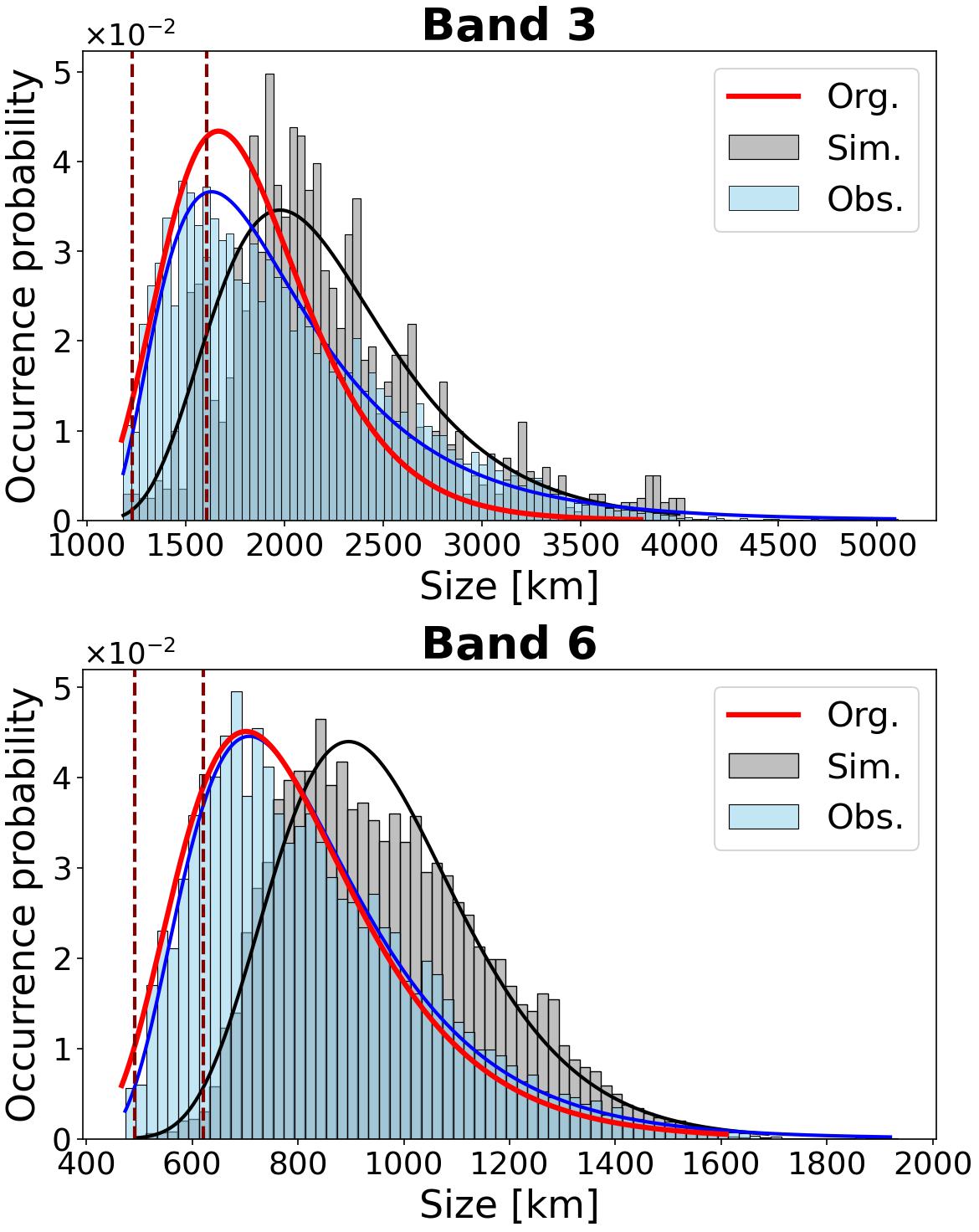}
    \caption{Distribution of sizes for all the features at all frames for Band 3 in the top and Band 6 in the bottom. In both panels the blue histograms correspond to results from observations and the grey to results from degraded simulations. The solid lines are their corresponding log-norm fits. {The red solid line corresponds to the log-norm fit of the results from the simulation at the original resolution.} The vertical maroon lines shows the minor and major beam axes. \label{fig:sizes}}
\end{figure}

The Band 3 observation consists of a time series split into 3 scans (blocks of observation), with a duration of about 10\,minutes each and a cadence of 2\,s. There are gaps of about 2 minutes between scans that ALMA uses for observing calibration sources. The pixel size chosen during the reconstruction process is 0.34\,arcsec. The ALMA synthetic elliptical beam size defines the interferometric spatial resolution of the observations and depends on the observed frequency, the time of the day, and the atmospheric conditions. For one specific frequency, the elliptical beam shape smoothly changes as time progress. In this observation lasting for about 35 minutes including calibration breaks, the median area of the synthetic elliptical beam was $2.94$ arcsec$^2$ with a standard deviation of $0.04$ arcsec$^2$ throughout the whole observation, and its dimensions were about $2.2$\,arcsec $\times$ $1.7$\,arcsec ($1600$\,km $\times$ $1235$\,km assuming that $1$ arcsec $\approx$ 727\,km) with a clockwise median inclination of $28.7^\circ$ with respect to the solar north.

The Band 6 observation consists of a times series with split into 5 scans, each with a cadence of 2\,s. The first 4 scans are $\sim$8 minutes long, with the final scan lasting $\sim$4 minutes. There are calibration gaps of about $\sim$2~minutes between scans. The pixel size chosen during the image reconstruction process was 0.14 arcsec. In this observation lasting about 45 minutes including breaks, the median area of the ALMA synthetic elliptical beam was $0.45$ arcsec$^2$ with a standard deviation of $0.03$ arcsec$^2$. The median shape of the beam was about $0.84$ arcsec $\times~0.67$ arcsec ($610$ km $\times$ $488$ km) with a clockwise median inclination of $86.1^\circ$ with respect to the solar north.

These ALMA observations have been spatially co-aligned with observations from the Solar Dynamic Observatory \citep[SDO;][]{2012SoPh..275....3P}, consisting of photospheric magnetograms from the Helioseismic and Magnetic Imager \citep[HMI;][]{2012SoPh..275..207S}. The solar coordinates of the centres of the field of views at the start of observations are $(x,y)=(-246,267)$ (arcsec) with a radius of 33 arcsec and $(x,y)=(-255,263)$ (arcsec) with a radius of 16 arcsec for bands 3 and 6 respectively. Both bands sample a plage region with strong magnetic fields spanning a range of \mbox{[-1084,59]}~Gauss in the photosphere. These data products are presented in \cite{2021RSPTA.37900174J} as data sets D2 and D8, where they showed that the magnetic topology at chromospheric heights represents vertical or near-vertical fields in a large portion of the FoVs with more inclined field towards the edges. This magnetic field topology verifies that these data sets are good candidates to search for MHD waves. In particular, the Band 3 observations have already proved to exhibit small-scale blob-like bright features with oscillations in temperature and size which reflect the characteristics of fast-sausage MHD-modes, with a transverse component likely being kink MHD-modes \citep{2021RSPTA.37900184G}. 

The left panels of Figure~\ref{fig:context_sdo} shows the contextual total power (TP) {or full-disk} maps for the Band 3 (top) \& 6 (bottom) observations, characterised by the bright signatures of an active region. The area of interest for this study is the plage region denoted by black dashed squares, with small black circles showing the FoV of the ALMA interferometric observation. The middle-left panels show the SDO/HMI images corresponding to the black dashed square in the TP maps at the start of the observation, with red circles denoting the Band 3 FoV. This SDO/HMI image reveals the magnetic nature of the underlying photosphere. The middle-right panels display a composite of the Atmospheric Imaging Assembly \citep[AIA;][]{2012SoPh..275...17L} at 170\,nm (pink) and 17.1\,nm (yellow), and shows the structuring of the coronal magnetic field. The right panels show Band 3 \& 6 snapshots with the beam shape visualised in the bottom-left corner by an ellipse. The white circle in the top panel illustrates the ALMA Band 6 FoV, with the green dots are the median location of all the small-scale bright features analysed in this study (see~Section~\ref{sec:analysis}). Finally, the white square with the blue cross in the Band 6 panel marks the position of a single feature which is used for illustrative purposes in this paper (see Fig.~\ref{fig:featexamp_snaps}).Through inspection of the images, it is clear that Band 6 can resolve magnetic structures with greater fidelity in comparison with Band 3. 

The specific time series used in this study can be downloaded from the Solar ALMA Science Archive (SALSA)\footnote{See \url{http://sdc.uio.no/salsa/}} as data sets D02 (Band 3) and D08 (Band 6) \citep[for further details on SALSA see][]{2022A&A...659A..31H}. The Solar ALMA Library of Auxiliary Tools (SALAT)\footnote{See \url{https://solaralma.github.io/SALAT/}} can be used for convenient data cube reading and loading of basic properties in header \citep[for more details on SALAT see][]{shahin_jafarzadeh_2021_5466873}.


\subsection{Simulation}
\label{sec:simulation}

The simulation used here was created with the 3D radiation magnetohydrodynamic (RMHD) code Bifrost \citep{2011A&A...531A.154G} and represents a magnetic enhanced network region surrounded by quiet Sun  \citep{2016A&A...585A...4C}. The simulation includes non-local thermodynamic equilibrium (non-LTE) radiative transfer and an equation of state that handles non-equilibrium hydrogen ionisation \citep{2015ApJ...802..136L}. The computational box extends from the upper convection zone to the corona with an extent of $24\text{ Mm} \times 24\text{ Mm} \times 16.8\text{ Mm}$ in the $(x,y,z)$ directions respectively. In the vertical direction $(z)$, the simulation uses an adaptive grid of 496 cells whose sizes vary between 19 km and 98 km, being of about 20 km at chromospheric heights, 14.4 Mm out of the 16.8 extent above the photosphere. For the horizontal directions $(x,y)$, the simulation has a grid of 504 cells with a constant size of 48 km, so that the pixel size of the simulation is equivalent to $0.066$ arcsec. 

The  simulation sequence used here has a cadence of 1\,s and lasts for about 1~hour of solar time, reaching an state of quasi-equilibrium after 200\,s \citep{2021RSPTA.37900185E}. The physical set-up of the simulation evolves in time with periodic boundary conditions in $(x,y)$ directions and open boundary conditions in the vertical direction. The magnetic field strength has an average unsigned value of 50\,G at photospheric heights. The initial condition is comprised of two photospheric regions of opposed polarity with a separation of 8\,Mm between them, which connect by magnetic loops in the atmosphere above \cite[for a detailed description of the simulation see][]{2016A&A...585A...4C}. This simulation is of interest to this study as MHD oscillations have been detected and analysed in fibrillar structures at coronal heights \citep{2021A&A...647A..81K}. This implies that MHD-modes may be present throughout the atmosphere, including the height range sampled by ALMA. 

We used the Advanced Radiative Transfer code \citep[ART;][]{2021_art} which solves the radiative transfer equation under the assumption of Local Thermodynamics Equilibrium (LTE). In this work, ART uses the non-equilibrium electron densities provided by the Bifrost simulations, as this is important for the resulting continuum radiation at sub-mm and mm wavelengths \cite[for a extended discussion of the use of simulations in the context of ALMA see][]{2022FrASS...9.7878W}. The final data product consists of two time series of continuum brightness temperatures at the frequencies of ALMA bands 3 and 6 (see Section~\ref{sec:observations}). In order to ensure a fair comparison between the observations and the simulation, we had to match the time resolution to those of ALMA. For this, we have arbitrary chosen the snapshot corresponding to 700 s as starting point\footnote{As a side note, we also did a test with a different starting snapshot and it did not change the final results} and so on removed every other frame, constructing a new time series wit the frames 700, 702, 704 and etc, until eventually we matched both ALMA observations time stamps. In the case of Band 3, the 3 scans are composed of the snapshots [700-1288, 1428-2022, 2172-2764]. For the case of Band 6, the 5 scans correspond to the snapshots [700-1182, 1294-1770, 1913-2389, 2499-2981, 3120-3354]. 

After matching the time resolution of ALMA, {we have also matched the spatial resolution so we can compare results between the original and the degraded (ALMA equivalent) resolution for each band.} The spatial resolution was amended through convolving each snapshot with the corresponding ALMA synthetic elliptical median beam to simulate ALMA's angular resolution. Please note that this approach assumes a perfect sampling of the spatial Fourier space while, in reality, ALMA's Fourier space coverage is sparse. Simulating these effects in detail, however, would go beyond the scope of the study presented here. Subsequently, the pixel size of the simulation was changed from 0.066 arcsec to the respective 0.34 arcsec in Band 3 and 0.14 arcsec in Band 6. The resulting time series of degraded mm brightness temperature maps {altogether with the equivalent original resolution time series are} the ones used for the analysis on this paper (see Sect.~\ref{sec:analysis}). Figure~\ref{fig:sim_convbeam} shows the same snapshot of the Bifrost simulation for the synthetic ALMA bands 3 and 6 in the top and bottom row respectively. The images in the left column correspond to the original simulation which roughly spans a range of temperatures between {2700\,K and 20000\,K} for Band~3 and {2750\,K and 16000\,K }for Band~6. The absolute temperature ranges of the degraded simulations are [3954,14280]\,K for Band~3 and [3005,12726]\,K for Band~6. {The mean temperature vales are the same in both cases, with 7337\,K for Band~3 and 5779\,K for Band~6.} In {all the simulations}, it is possible to visually identify small structures and the typical fibrillar nature of the quiescent chromosphere.The images in the right column correspond to the simulation after convolving with the ALMA beams and resizing to the pixel size of the ALMA data{, i.e., the so-called degraded simulation}. The small white ellipses in the bottom-left corners represent the interferometric beams. The most evident effect of the convolution is the degradation of the spatial resolution, which blurs the structures in the simulation, giving them a more blob-like shape. The convolution also has an impact on the temperatures, reducing the range of observed values, which has to be taken into account when searching for temperature enhancements. The green dots {in all the panels} mark the median location of all the small-scale bright features detected in the simulations and the blue cross in the Band 6 {degraded} snapshot mark the location of a feature which has been used to illustrate the analysis (see Figure~\ref{fig:featexam_cwt}). In addition, it is worthwhile noting that the area of the FoV in the Band 3 observation is roughly 4 times the area of the FoV in the Band 3 simulation. A complete description of the impact of the different ALMA observing modes in this Bifrost simulation can be read in \citet{2021A&A...656A..68E}.

\section{Feature Identification and Tracking}
\label{sec:methods}

\begin{figure*}[h]
    \sidecaption
    \includegraphics[width=.85\textwidth]{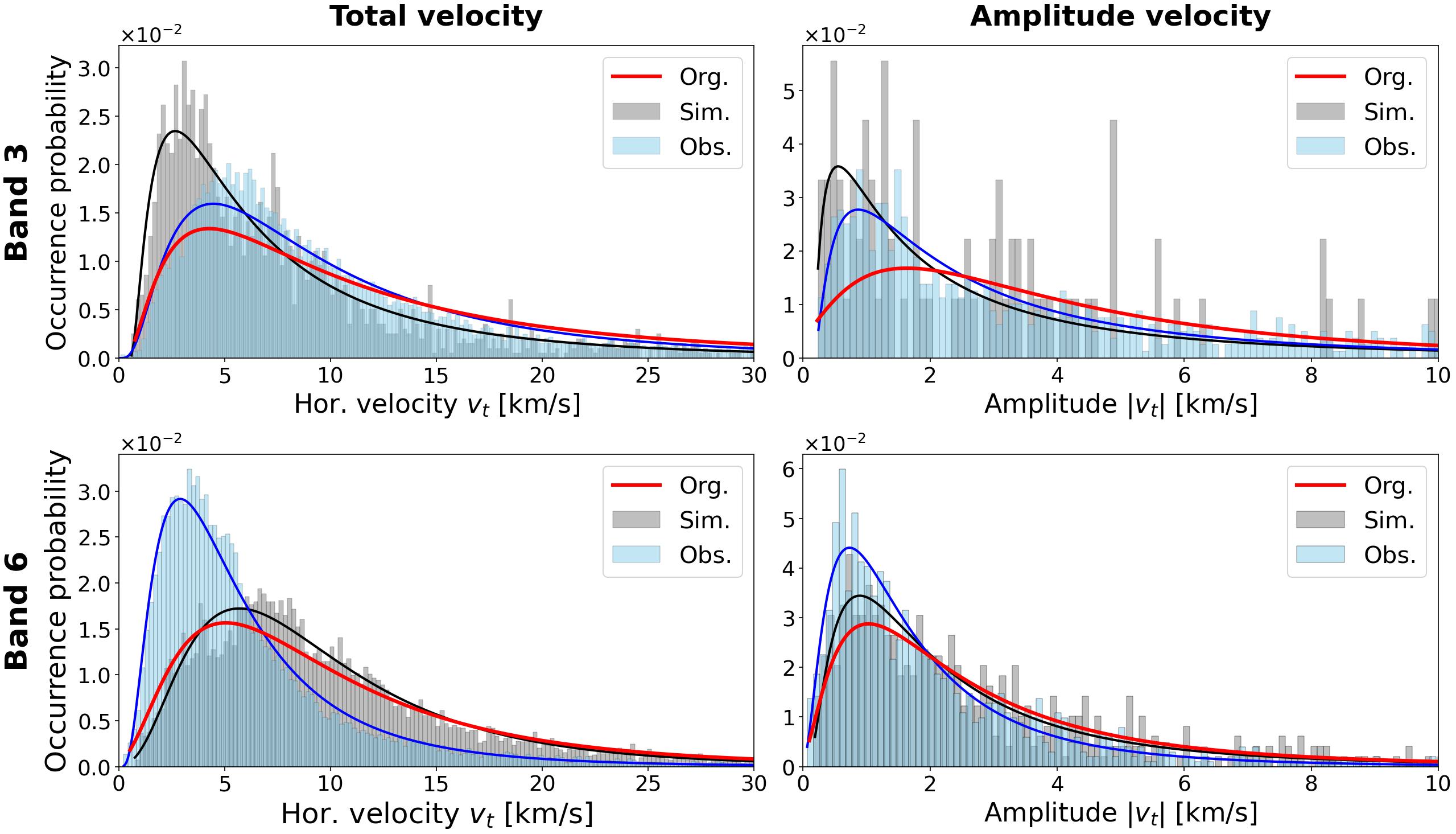}
    \caption{Distributions of total horizontal velocities (left column) and amplitude velocities (right column) for {degraded} simulations (grey) and observations (blue) in the features found and traced in Bands 3 (top row) and 6 (bottom row). The blue and black solid lines correspond to a log-norm fit in each corresponding case. {The red solid lines correspond to the log-norm fits of the results from the simulation at the original resolution.}\label{fig:velocities}}
\end{figure*}

A visual inspection of the ALMA observations reveals the presence of transient events in the shape of small-scale bright features  that appear to move while their brightness and size vary (see movies in SALSA). Moreover, \citet{2021RSPTA.37900184G} showed by means of wavelet transform analysis that three of these small-scale features found in the Band~3 observation were likely of magnetic nature,  indicating that their oscillatory behaviour are signatures of MHD-modes. However, these isolated cases were not sufficient to make a statistically significant conclusions. Accordingly, in this study we aimed to analyse the oscillatory behaviour of as many features as possible, in both the two observed ALMA bands, (Sect.~\ref{sec:observations}) and the two synthetic continuum map time series from the Bifrost simulation (Sect.~\ref{sec:simulation}). Therefore, it was necessary to implement an algorithm to automatically identify and track the features across the time series such that the output from this algorithm could be used to perform a statistical analysis.

Identifying and tracking varying features in a series of ALMA images is not a minor task, as the chromosphere contains a variety of features, each with different sizes, shapes and intensities. These structures can commonly interact and overlap within the FoV, resulting in extra-ordinary shifts and morphological changes from one frame to the next one. To constrain the problem, it was imperative to establish some threshold characteristics that any of the bright features had to meet in order to be selected by the algorithm. In particular, the feature size was the main property taken into account. By manual inspection, we found that the sizes of the features were around the sizes of the corresponding ALMA elliptical beams (i.e., the spatial-resolution proxy). This fact was in agreement with the previous results of \citet{2021RSPTA.37900184G},  with sizes between 0.95 and 1.27 times the median beam's major axis.

\begin{table*}[tp!]
\centering
\begin{tabular}{cc|ccc|ccc|}
\cline{3-8}
\multicolumn{1}{l}{} &
  \multicolumn{1}{l|}{\textbf{}} &
  \multicolumn{3}{c|}{\textbf{Band 3}} &
  \multicolumn{3}{c|}{\textbf{Band 6}} \\ \cline{3-8} 
\textbf{} &
   &
  \multicolumn{1}{c|}{\textbf{ALMA Obs.}} &
  \multicolumn{1}{c|}{\textbf{Deg. Sim.}} &
  \textbf{Org. Sim} &
  \multicolumn{1}{c|}{\textbf{ALMA Obs.}} &
  \multicolumn{1}{c|}{\textbf{Deg. Sim.}} &
  \textbf{Org. Sim.} \\ \hline
\multicolumn{1}{|c|}{} &
  \textbf{\# Feat.} &
  \multicolumn{1}{c|}{193} &
  \multicolumn{1}{c|}{24} &
  36 &
  \multicolumn{1}{c|}{293} &
  \multicolumn{1}{c|}{204} &
  392 \\ \cline{2-8} 
\multicolumn{1}{|c|}{} &
  \textbf{$\frac{\# Feat.}{Mm^2 s}$} &
  \multicolumn{1}{c|}{$6.1\times 10^-5$} &
  \multicolumn{1}{c|}{$2.3\times 10^-5$} &
  $3.4\times 10^-5$ &
  \multicolumn{1}{c|}{$3.2\times 10^-4$} &
  \multicolumn{1}{c|}{$1.6\times 10^-4$} &
  $3.1\times 10^-4$ \\ \hline
\multicolumn{1}{|c|}{\multirow{3}{*}{\textbf{\begin{tabular}[c]{@{}c@{}}Hor. Amplitude\\ velocity {[}\,km\,s$^{-1}${]}\end{tabular}}}} &
  \textbf{Range} &
  \multicolumn{1}{c|}{0.2-33.8} &
  \multicolumn{1}{c|}{0.2-38.9} &
  0.2-37.2 &
  \multicolumn{1}{c|}{0.1-12.3} &
  \multicolumn{1}{c|}{0.2-17.3} &
  0.1-18.2 \\ \cline{2-8} 
\multicolumn{1}{|c|}{} &
  \textbf{Mean} &
  \multicolumn{1}{c|}{4.2} &
  \multicolumn{1}{c|}{4.1} &
  5.0 &
  \multicolumn{1}{c|}{2.1} &
  \multicolumn{1}{c|}{2.8} &
  3.1 \\ \cline{2-8} 
\multicolumn{1}{|c|}{} &
  \textbf{Median} &
  \multicolumn{1}{c|}{2.6} &
  \multicolumn{1}{c|}{2.6} &
  3.7 &
  \multicolumn{1}{c|}{1.5} &
  \multicolumn{1}{c|}{2.1} &
  2.3 \\ \hline
\multicolumn{1}{|c|}{\multirow{3}{*}{\textbf{\begin{tabular}[c]{@{}c@{}}Geometric\\  heights {[}km{]} $\tau=1$\end{tabular}}}} &
  \textbf{Range} &
  \multicolumn{1}{c|}{-} &
  \multicolumn{1}{c|}{1152-2269} &
  725-3278 &
  \multicolumn{1}{c|}{-} &
  \multicolumn{1}{c|}{757-2368} &
  506-3081 \\ \cline{2-8} 
\multicolumn{1}{|c|}{} &
  \textbf{Median} &
  \multicolumn{1}{c|}{-} &
  \multicolumn{1}{c|}{1563} &
  1573 &
  \multicolumn{1}{c|}{-} &
  \multicolumn{1}{c|}{1260} &
  1387 \\ \cline{2-8} 
\multicolumn{1}{|c|}{} &
  \textbf{Std. Dev.} &
  \multicolumn{1}{c|}{-} &
  \multicolumn{1}{c|}{192} &
  295 &
  \multicolumn{1}{c|}{-} &
  \multicolumn{1}{c|}{171} &
  201 \\ \hline
\multicolumn{1}{|c|}{\multirow{3}{*}{\textbf{Feature Sizes {[}km{]}}}} &
  \textbf{Range} &
  \multicolumn{1}{c|}{1183-5088} &
  \multicolumn{1}{c|}{1183-3995} &
  1176-3804 &
  \multicolumn{1}{c|}{473-1920} &
  \multicolumn{1}{c|}{493-1688} &
  467-1606 \\ \cline{2-8} 
\multicolumn{1}{|c|}{} &
  \textbf{Median} &
  \multicolumn{1}{c|}{1895} &
  \multicolumn{1}{c|}{2116} &
  1819 &
  \multicolumn{1}{c|}{789} &
  \multicolumn{1}{c|}{943} &
  777 \\ \cline{2-8} 
\multicolumn{1}{|c|}{} &
  \textbf{Std. Dev.} &
  \multicolumn{1}{c|}{583} &
  \multicolumn{1}{c|}{535} &
  403 &
  \multicolumn{1}{c|}{218} &
  \multicolumn{1}{c|}{195} &
  201 \\ \hline
\multicolumn{1}{|c|}{\multirow{2}{*}{\textbf{\begin{tabular}[c]{@{}c@{}}Representative\\ T$_b$ {[}K{]}\end{tabular}}}} &
  \textbf{Median} &
  \multicolumn{1}{c|}{11083} &
  \multicolumn{1}{c|}{9789} &
  10970 &
  \multicolumn{1}{c|}{9000} &
  \multicolumn{1}{c|}{7136} &
  8621 \\ \cline{2-8} 
\multicolumn{1}{|c|}{} &
  \textbf{Std. Dev.} &
  \multicolumn{1}{c|}{949} &
  \multicolumn{1}{c|}{1287} &
  1432 &
  \multicolumn{1}{c|}{671} &
  \multicolumn{1}{c|}{874} &
  1006 \\ \hline
\end{tabular}
\caption{{Number of features, density of features per square mega-meter per second, amplitudes of horizontal velocities, heights of optical depth unity ($\tau=1$, for features found in the simulation), feature  sizes and representative brightness temperatures.} \label{tab:basic_qtt_v2}}
\end{table*}

With a size threshold, and following the essence of the method of \citet{1996JCIS..179..298C} to identify, track and extract quantitative information from colloidal suspensions in noisy images (implemented to small-scale solar magnetic elements by \citealt{2013A&A...549A.116J}), we applied to each frame a Difference of Gaussians (DoG) filter with Gaussian widths of half and double the median major axis of the ALMA beam. This filter enhances the blob-like features making their detection easier (see Appendix~\ref{apx:A}). Then, for each frame, we saved the position of all local maxima separated by at least 1.5 times the median major beam. Subsequently, we used the positions of the local maxima as seed points in the Flood fill method implemented in the Python \texttt{Scikit-image} library \citep{scikit-image}. The flooding method is applied to a normalised, inverted colour scale of the ALMA images, i.e., those where bright pixels were made dark and vice-versa. The flooding algorithm begins at a starting point or seed position (each local maximum) and then checks if the pixel values of surrounding positions fall within the so-called tolerance range with respect to the value of the seed point. This range was chosen such that for each seed point the method would flood all surrounding pixels within $\sim2.4$  standard deviations, corresponding to the full width at half maximum {understood as all the pixels where the temperature is halfway between the pixel seed position (local maximum) temperature and the median temperature of the complete FoV}. This yields a mask of flooded pixels for each frame, in which it is possible to identify the isolated features as small flooded regions completely surrounded by non-flooded pixels. As we are interested in blob-like features of sizes similar to the beam size, all the features with distances between any two points on their borders larger than 4 times the median beam major axis were discarded. This step serves as a filter for large features, but also for features with elongated shapes. This process yields a cube of masked images in which every individual mask represent a possible feature of interest. To track the features in time, the overlapping of the masks is compared frame by frame{, meaning that if the mask seen at the time $t_{i-1}$ shares at least 2 pixels with the mask seen at the time $t_i$ then they are considered as belonging to the same feature. In the case that two or more masks merge in the following frame, each individual feature is considered alive until the previous frame. When all the features are individualised, each of those obtains and obtains an unique identifier.} {This} process results in a cube that contains the masks of all individual features and their unique labels. 
Such a cube is produced for both the observational and simulated ALMA data for both bands and has the same dimensions as the respective original data cube. {Finally, the maximum brightness temperature of each unique feature in each frame is extracted and used to construct light-curves for each feature. These light curves are classified by means of the Density-Based Spatial Clustering of Applications with Noise (DBSCAN) algorithm which is part of the Python Machine Learning Scikit-learn package \citep{scikit-learn}. This classification results in different clusters of light-curves, namely, oscillating light-curves (see e.g., Fig.~\ref{fig:featexam_cwt}), light-curves with a sole transient brightening \cite[see e.g. Fig.~1 at][]{2020A&A...644A.152E} and light-curves with any other different behaviour. We have then chosen all the features belonging to the cluster of light-curves with oscillatory behaviour to be analysed in this study.}


\section{Analysis and results}
\label{sec:analysis}

\subsection{Occurrence and general properties} 

The method described in Section~\ref{sec:methods} was applied to the time series of ALMA observations and simulations in both bands. In total, {193} features where identified in Band 3 and {293} in the Band 6 observations. {In the simulation at the degraded resolution 24 and 204 features were identified in bands 3 and 6, respectively. While in the original resolution 36 and 392 features in bands 3 and 6 were identified,  respectively.} The raw number of features does not tell anything about how similar the results in the observation  are compared to the simulation. Thus, to be able to compare the results, it is necessary to take into account the duration of each time series and the area of the FoVs. The resulting number density values, which are defined as the total number of features per square mega-metre per second, are listed in Table~\ref{tab:basic_qtt_v2}. These density values fall within the same order of magnitude for the observation and simulation in each band. However, the densities in the observational data are approximately {double for Band 6 and triple for Band 3 than the densities found in the degraded simulation for those bands. While the density values found at the simulation with original resolution are closer to those in the observation, specially for Band 6. This is a direct consequence of the different spatial resolutions at each band. An additional} possible explanation for at least part of this systematic difference  may be due to the 
differences in magnetic topology of the observed and simulated field. The FoV of the observations is mostly covered by the magnetic enhanced region with long fibrillar structures (see Fig.~\ref{fig:context_sdo}), whereas in the simulation the magnetic enhanced region is comparatively smaller and confined to the central FoV region only (see Fig.~\ref{fig:sim_convbeam}). Corresponding differences in the occurrence of transient bright features as analysed here are then expected.  
We also note that the events are a factor 5-6 more frequent in Band~6 than in Band~3, which is seen in both the observations and simulations. The higher angular resolution of Band~6 compared to Band~3 (a factor of 2.3) may play a part, allied with the height differentials of the atmospheric layers mapped by Band~3 and Band~6 (see Sect.~\ref{sec:discussion}).

For each individual feature, the algorithm returns its shape, area, value of the brightest pixel, average pixel value within the feature, initial and final time, and the position of the centre of gravity (of intensity) as computed from the pixel values within the feature's border as weights. For the purpose of this paper, all the areas were converted to the diameter of a circle of equal area. This equivalent diameter is hereafter referred to as the size of the feature.The representative brightness temperature of the features has been chosen to be the value of the brightest pixel with median and standard deviation values in Table~\ref{tab:basic_qtt_v2}.

\begin{figure}[tp!]
    \centering
    \includegraphics[width=.45\textwidth]{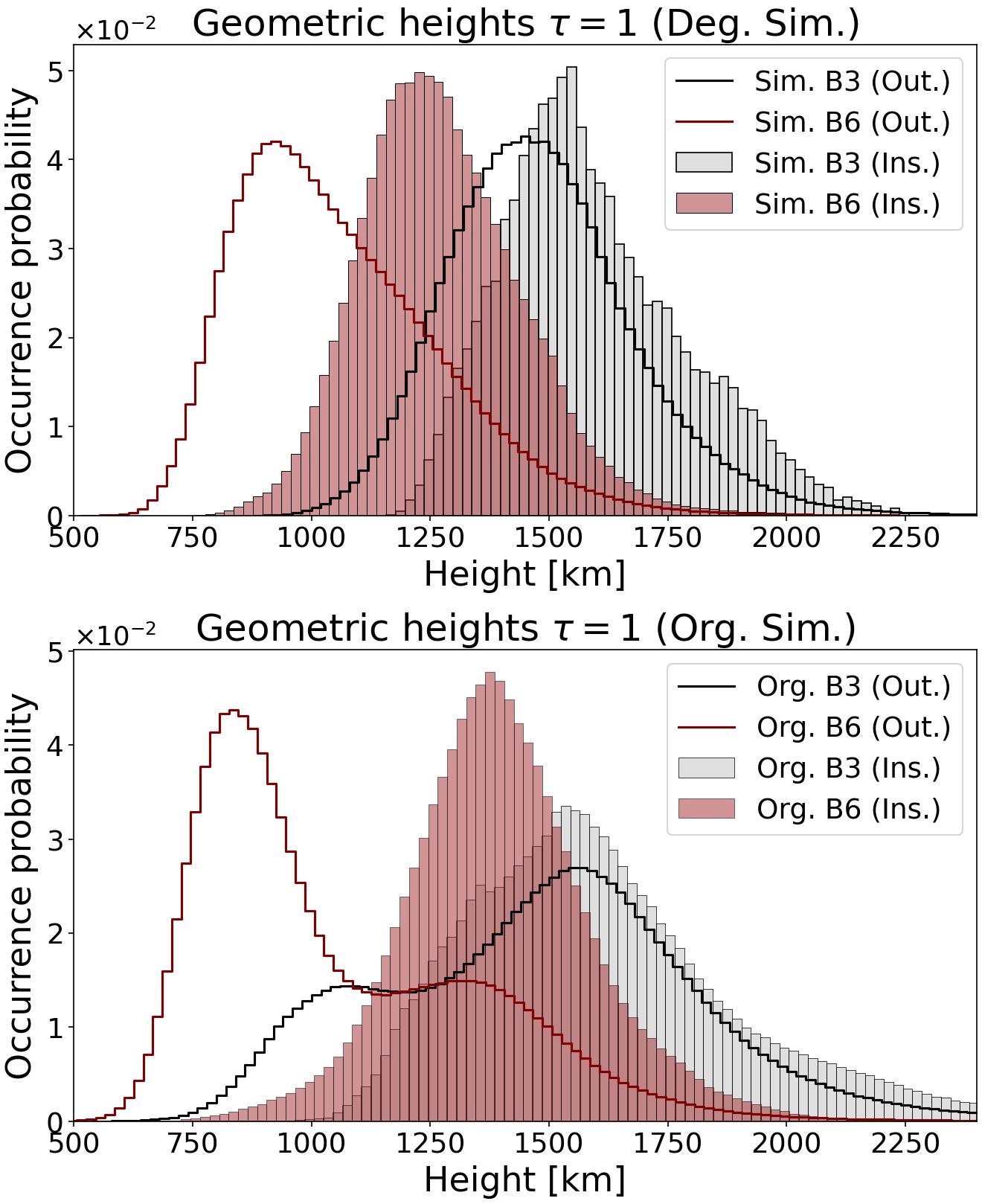}
    \caption{The distribution of geometric heights from {the original (top panel) and degraded (bottom panel) simulations }inside and outside the feature borders. The grey-black colours correspond to values in Band 3 and the red colours to values in Band 6. \label{fig:heights}}
\end{figure}

\begin{figure}[tp!]
    \includegraphics[width=.48\textwidth]{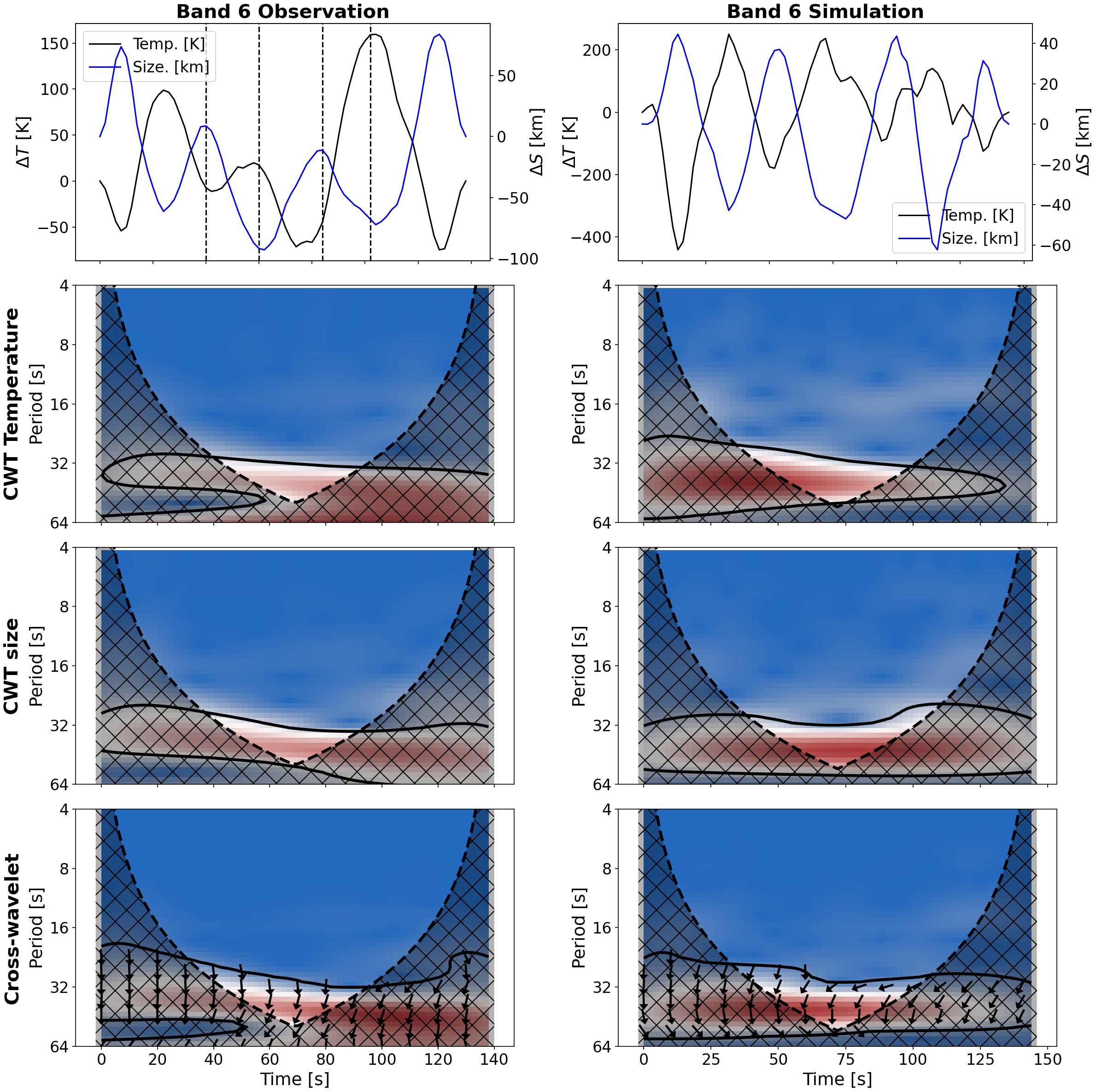}
    \caption{Temperature and size oscillations of an example feature in Band 6 observation (left column) and {degraded}simulation (right column) are shown in the top row. From the second row to bottom the continuous wavelets for temperature, size and the cross-wavelet for the same two quantities are shown. For these panels the period is plotted using a log$_2$ scale in the y-axis and time is shown in x-axis. The hashed black region highlights the cone of influence for each spectrum and the black solid lines mark the 95\% confidence levels. The arrows in the bottom panels depict the phase difference between temperature and size oscillations, being in-phase when arrows point straight up, anti-phase when arrows point straight down, pointing to the left indicate that perturbation of temperature follows those of size and vice versa when pointing to the right. \label{fig:featexam_cwt}}
\end{figure}

\begin{figure*}[tp!]
    \centering
    \includegraphics[width=.95\textwidth]{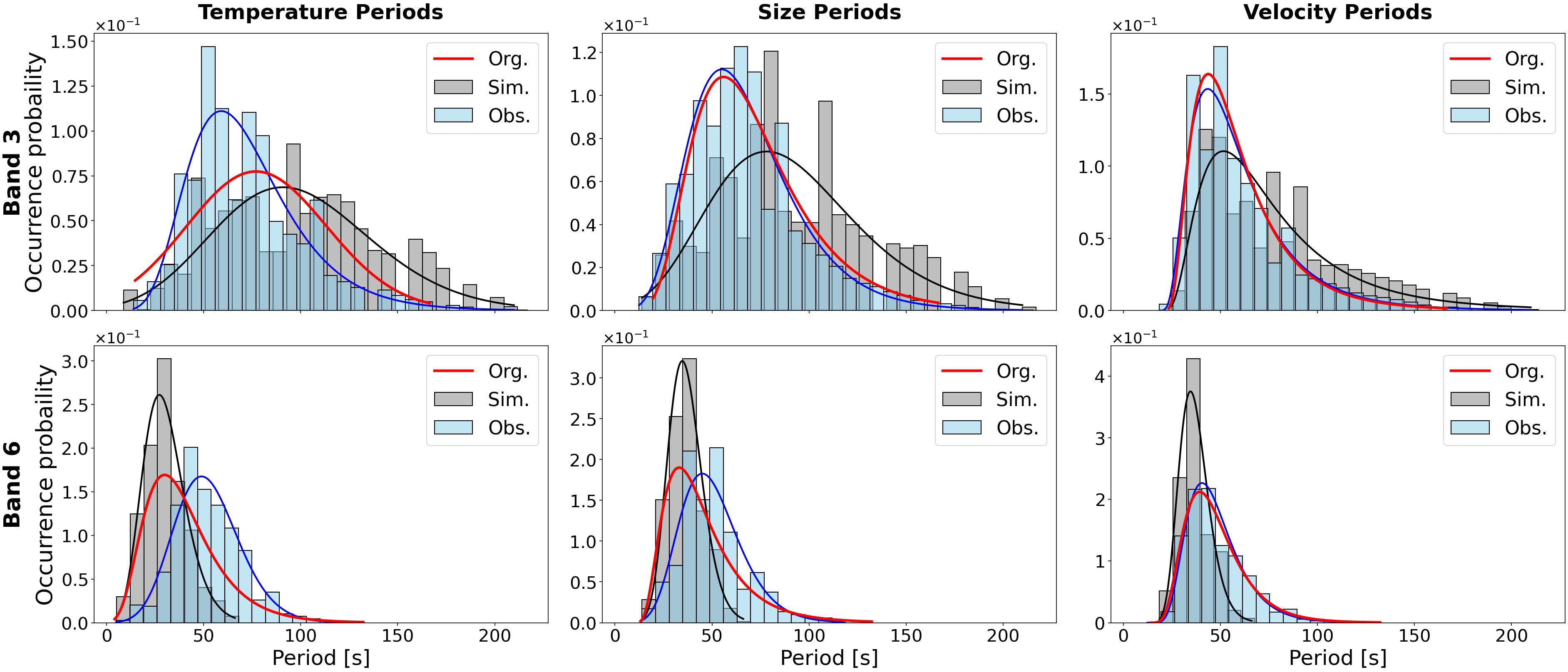}
    \caption{Distribution of dominant periods for Band 3 in the top row and Band 6 in the bottom row, for temperature (left column), size (middle column) and horizontal velocity (right column). The degraded simulation's distributions are shown in grey, the observation's distributions in blue{ and the original simulation's distribution is show as a red line.}\label{fig:periods}}    
\end{figure*}

\begin{table*}[ht]
\centering
\begin{tabular}{@{}l|ccc|ccc|ccc|@{}} \cline{2-10}
\multicolumn{1}{c|}{}                  &  \multicolumn{3}{c|}{\textbf{Temperature periods {[}s{]}}} & \multicolumn{3}{c|}{\textbf{Size periods {[}s{]}}}   & \multicolumn{3}{c|}{\textbf{Velocity periods {[}s{]}}} \\ \cline{2-10}
\multicolumn{1}{c|}{}                  & \textbf{Mean}   & \textbf{Median}   & \textbf{Std. Dev.}  & \textbf{Mean} & \textbf{Median} & \textbf{Std. Dev.} & \textbf{Mean}  & \textbf{Median}  & \textbf{Std. Dev.} \\ \hline
\multicolumn{1}{|l|}{\textbf{ALMA B3}} & 72              & 66                & 29                  & 68            & 62              & 28                 & 61             & 52               & 27                 \\ \hline
\multicolumn{1}{|l|}{\textbf{Sim. B3}} & 99              & 99                & 41                  & 92            & 83              & 40                 & 78             & 70               & 36                 \\ \hline
\multicolumn{1}{|l|}{\textbf{Org. B3}} & 78              & 83                & 36                  & 72            & 66              & 30                 & 60             & 52               & 25                 \\ \hline
\multicolumn{1}{|l|}{\textbf{ALMA B6}} & 52              & 50                & 17                  & 50            & 50              & 16                 & 48             & 44               & 15                 \\ \hline
\multicolumn{1}{|l|}{\textbf{Sim. B6}} & 30              & 29                & 11                  & 37            & 35              & 9                  & 37             & 37               & 8                 \\ \hline
\multicolumn{1}{|l|}{\textbf{Org. B6}} & 40              & 37                & 20                  & 45            & 42              & 20                 & 48             & 44               & 17                 \\ \hline
\end{tabular}
\caption{{The mean, median and standard deviation of oscillation periods for temperature, size and horizontal velocity}\label{tab:period}}
\end{table*}

\subsection{Location{, lifetimes,} size and horizontal velocity}
\label{sec:loc_size_vel}

Figure~\ref{fig:featexamp_snaps} illustrates the outcome of the method previously described. From left to right it shows the time evolution of one of the many features identified and traced in Band 6 which has been highlighted in the bottom-right panel of Figure~\ref{fig:context_sdo} as a blue cross in the middle of a white square. In Fig.~\ref{fig:featexamp_snaps}, the range of the colour map is uniform throughout the 4 panels which correspond to the times indicated as vertical lines in the top left panel of Fig.~\ref{fig:featexam_cwt}. The black contours correspond to the borders of the feature. Although the shape of the feature changes with time, it remains close to a blob-like shape, making the diameter of a circle of equal area an excellent proxy for its size and its corresponding evolution. The rightmost panel shows the displacement of the centre of gravity during the lifespan of the feature, while the arrow shows the direction of time. Each displacement of the centre of gravity in $x$ and $y$ directions between 2 consecutive frames was used to calculate the instantaneous horizontal velocities $v_x$ and $v_y$ and the corresponding  instantaneous horizontal velocity $v_t = \sqrt{v_x^2 + v_y^2}$ for each feature. 

{Figure~\ref{fig:Lifes} shows the distribution of lifetimes for all features found in the observation in blue, features found in the degraded simulation in grey and features found in the original simulation as a solid red line. Band 3 is shown in the top panel and band 6 in the bottom panel, in both bands, the peaks of the distributions are between 80 and 120 seconds. The lifetimes of the observations are a bit shifted to the right compared with the lifetimes of the simulation.}

Table~\ref{tab:basic_qtt_v2} summarises the range, median and standard deviation of the sizes and Figure~\ref{fig:sizes} shows the distributions of sizes of all the features at all time frames in observations and simulations. Band~3 is shown in the top panel and Band~6 in the bottom panel. In both panels, the vertical maroon dashed lines are the median minor and major axes of the ALMA beam in each corresponding band, showing that the features found by the method were completely resolved by ALMA. The grey distributions correspond to results from the{ degraded }simulations and the blue distributions to results from observations. The blue and black thick lines shows a log-norm fit in each case. {The red solid lines show the log-norm fit of the results from the original resolution simulation}. The peaks of the distributions are located at larger sizes in {degraded }simulations compared to observations although the shapes of the distributions are alike. {The distributions of sizes in the case of the original simulation are very similar to the distributions found in the observations. This may be due to the fact that the degrading of the resolution results in a wider FWHM as the median temperature remains the same but the higher temperatures are diminished.} However, as it will be discussed in Sect.~\ref{sec:discussion}, the results between observations and simulations do not differ significantly, implying that the difference in the peaks may be due just to a discrepancy in the spatial power spectrum between observations and simulation given by the apparently more grainy pattern present in the observations. This is likely  caused to some extent by ALMA's Fourier space coverage of the observations being  sparse whereas perfect Fourier space coverage is assumed for the simulations (see also Sect.~\ref{sec:simulation}). 

Figure~\ref{fig:velocities} shows the distribution of the horizontal velocities for Band 3 in the top row and for Band 6 in the bottom row. Again, the grey distributions correspond to results from simulations, the blue distributions to results from observations and the blue and black thick lines correspond to log-norm fits. The distributions of horizontal velocities are shown in the left column, while the distributions of the velocity amplitudes  are shown in the right column. The velocity amplitudes  are computed by taking the fast Fourier transform of each horizontal velocity oscillation and subsequently choosing those amplitude values greater than two standard deviations over the average. Table~\ref{tab:basic_qtt_v2} summarises the ranges, means and median velocity amplitudes in the four cases. 
 
Figure~\ref{fig:heights} shows the distributions of formation heights at $\tau=1$ in each band {from the degraded simulation at the top panel and the original simulation at the bottom panel.} The shadow regions represent the heights inside the features while the the solid lines show the heights outside the features. Interestingly, the radiation is formed higher in the solar atmosphere within the features {both in the degraded and original simulations. This higher formation is more noticeable in Band 6 than in Band 3.} This fact is in agreement with the behaviour found by \cite{2015A&A...575A..15L} \& \cite{2021A&A...656A..68E} in the same simulation. In particular, the radiation formed above regions of stronger magnetic field at ALMA wavelengths comes from higher heights when compared to the radiation formed above more quiet regions.

\subsection{Wavelet analysis and periods}
\label{subsec:wavelet_per}

The time evolution of the features temperature, size and
horizontal velocity exhibits an oscillatory behaviour. 
In order to determine the  oscillation periods, we performed a continuous wavelet analysis (CWT) for temperature, size and velocity for each individual feature. This allows for the analysis of a time-varying signal in both frequency and time domains by determining the dominant modes of oscillation in the power spectrum. A detailed mathematical description of the method can be found in \citet{1998BAMS...79...61T}. In addition, it is also possible to perform a cross-wavelet analysis to identify the phase difference between two signals. The cross-wavelet is useful when we know that the variation of two signals is linked, for example, due to physical reasons. Wavelet analysis has been widely used to characterise oscillations in the solar atmosphere \citep[see, e.g.,][and reference there in]{2004ApJ...604..936B,2007A&A...473..943J,2017ApJS..229....9J,2021RSPTA.37900184G}

To proceed with the analysis of these oscillations, all time series corresponding to temperature, size and horizontal velocity were detrended (using a linear fit) and apodised (using a Tukey window with a length of 0.1). Figure~\ref{fig:featexam_cwt} shows the temperature (black lines) and size (blue lines) oscillations of two features in Band 6 in the top row. The Band 6 detection (left column) corresponds to the same feature shown in Fig.~\ref{fig:featexamp_snaps},  with the {degraded} simulation detection (right column) exhibiting a similar morphology, highlighting the complementary nature of features detected in observations and simulation. The continuous wavelet transforms for temperature, size and cross-wavelet are shown, where black contours highlight the power regions above a 95\% confidence level for further analysis. For these two particular features, the dominant periods for both temperature and size are around 32 seconds. The cross-wavelet plots also show small arrows within the confidence level. These arrows represent the phase lag of the two signals, being in phase when they point downwards and anti-phase when they point upwards, visualising the phase difference between size and temperature oscillations at the dominant periods within the confidence level. The horizontal velocity oscillations and their continuous wavelet transforms look similar to those already reported in Figs.~5 and 8 of \citet{2021RSPTA.37900184G} {where the velocities for three features in the Band~3 observation reach values up to 20\,km\,s$^{-1}$ and average oscillating periods between 43 and 72\,s. Consequently, the plots regarding the velocities are not shown explicitly here.} 

The wavelet analysis described above is applied to each individual feature, resulting in the dominant  oscillation periods (i.e., all periods corresponding to wavelet areas with a confidence level of 95\% or larger and outside the cone of influence) as well as the phase differences between temperature and sizes. Table~\ref{tab:period} summarises the mean, median and standard deviation of the dominant periods from the wavelet analysis. Figure~\ref{fig:periods} shows from left to right the period distributions {of temperature, size and velocity for simulations in grey and solid red line, and observations in blue.} The distributions for Band 6 are narrower than for Band 3, with smaller periods, indicating that in general features oscillate faster at the chromospheric heights mapped by Band 6 as compared to the heights for Band 3.

\begin{table*}[]
\centering
\begin{tabular}{|l|c|c|c|c|c|c|}
\hline
\multicolumn{1}{|c|}{\textbf{Parameters}}                    & \textbf{ALMA B3}   & \textbf{Sim. B3}  & \textbf{Org. B3}   & \textbf{ALMA B6}   & \textbf{Sim. B6}  & \textbf{Org. B6}  \\ \hline
{Filling factor {[}\%{]}}                                    & 3.7 $\pm$ 0.8      & 1.6 $\pm$ 1.0     & 1.3 $\pm$ 0.8      & 2.7 $\pm$ 0.7      & 1.2 $\pm$ 0.5     & 1.9 $\pm$ 0.5     \\ \hline
{Temperature perturbation {[}\%{]}}                          & 1.1 $\pm$ 0.6      & 1.3 $\pm$ 0.8     & 1.2 $\pm$ 0.6      & 1.4 $\pm$ 0.8      & 1.9 $\pm$ 1.2     & 2.0 $\pm$ 1.1     \\ \hline
{Area perturbation {[}\%{]}}                                 & 14 $\pm$ 11        & 10 $\pm$ 6        & 14 $\pm$ 9         & 14 $\pm$ 9         & 12 $\pm$ 7        & 15 $\pm$ 10       \\ \hline
{Phase speed $\mathrm{v_{ph}}$ {[}\,km\,s$^{-1}${]}}                   & 14.7 $\pm$ 0.9     & 13.0 $\pm$ 1.0    & 13.0 $\pm$ 1.4     & 13.2$\pm$ 0.9      & 11.6 $\pm$ 1.0    & 11.6 $\pm$ 1.3    \\ \hline
{Radial velocity $\langle\mathrm{v_r}\rangle$ {[}\,km\,s$^{-1}${]}}    & 11.3 $\pm$ 0.3     & 9.8 $\pm$ 0.9     & 8.1 $\pm$ 0.5      & 5.6 $\pm$ 0.1      & 6.8 $\pm$ 0.1     & 7.2 $\pm$ 0.1     \\ \hline
{Energy flux {[}\,W\,m$^{-2}${]}}                                  & 1838 $\pm$ 429     & 621$\pm$ 407      & 453$\pm$ 289       & 5341 $\pm$ 1448    & 3640 $\pm$ 1557   & 5485 $\pm$ 1587   \\ \hline
\end{tabular}
\caption{Properties of the sausage modes oscillations. \label{tab:computed_par}}
\end{table*}

\subsection{Phase relations}
\label{subsec:phases_densities}

\begin{figure}[tp!]
    \sidecaption
    \includegraphics[width=.47\textwidth]{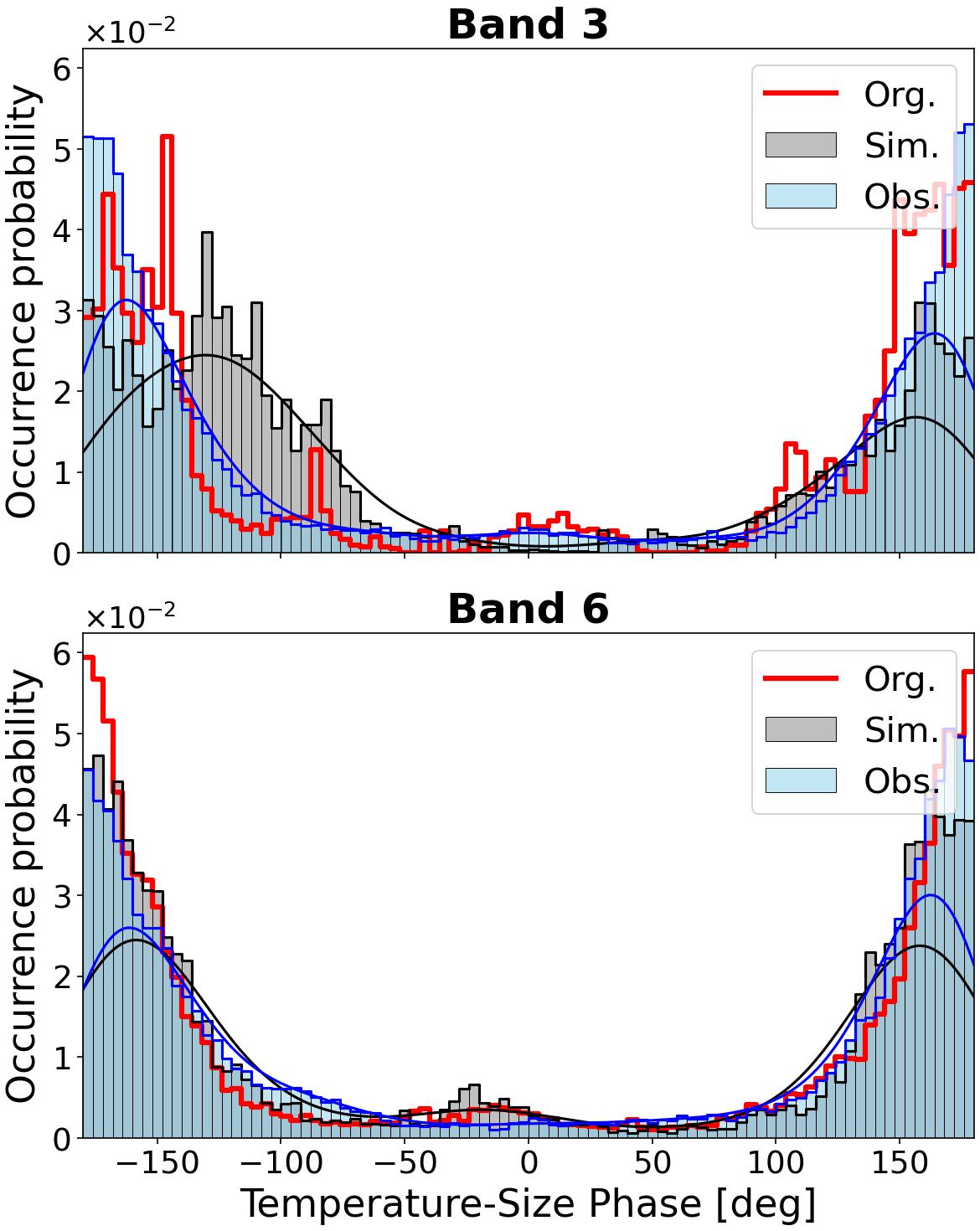}
    \caption{The distribution of phase angles between temperature and size for {degraded simulation (grey), observation (blue) and original simulation (red)} in ALMA Bands 3 (top) and 6 (bottom).}
    \label{fig:phases}
\end{figure}

Figure~\ref{fig:phases} shows the distribution of phase differences between temperature and size for all the features in Band 3 (top) and Band 6 (bottom). In both bands, the distributions for the observations (blue-shaded histograms) clearly exhibit a predominant anti-phase relation (-180$^{\circ}$ or 180$^{\circ}$) between size and temperature for both bands. For the {degraded} simulation (grey-shaded histograms), there is also a noticeable number of events with anti-phase relations. However, in Band 3 the distribution also shows substantial peaks at phases around{ -130$^{\circ}$ while in Band 6 this behaviour is not presented.} {The red solid lines shows the distributions for the features found in the original simulation. In the case of Band 3, the anti-phase relation is more clear than in the degraded simulation. Please note that the original simulation does show a peak around -90$^{\circ}$ which is likely enhanced and extended when the convolution with the ALMA beam is applied. In addition, the distributions for band 3 and 6 of original simulations show a small bump around 0$^{\circ}$.} This `in-phase' behaviour is also seen in the observations, but their relevance compared to the anti-phase behaviour is minimal. The phase relationships between structural properties have been used to derive the mode of the MHD wave \citep{2013A&A...555A..75M}, and would indicate that {a higher} combination of slow and fast sausage modes were present in the simulation {compared to the observation}. However these models depend on density and structural uniformity seen only in the photosphere, and have been shown to no longer be valid in the chromosphere \citep{2022ApJ...938..143G}. Instead, they suggest that that the physics present in the simulations, although similar to the physics of the observed plage region, cannot completely account for this particular portion of the real Sun.

\subsection{Energy flux estimation}
\label{subsec:wave_energy}

The time averaged flux of energy $\langle F \rangle$ carried by a wave through a surface perpendicular to a magnetic flux tube can be estimated using the following equation originally provided by \citet{2015A&A...578A..60M}:

\begin{equation}
    \langle F \rangle = \frac{1}{2} \, f \, (1+\ln(1/f)) \, \rho_e \, \mathrm{\langle v_r\rangle^2} \, \mathrm{v_{ph}} \ ,
    \label{equ:energy_flux}
\end{equation}

\noindent where $f$ is the filling factor, $\rho_e$ is the mass density outside the flux tube, $\mathrm{v_r}$ is the radial velocity of the perturbation and $\mathrm{v_{ph}}$ is the phase speed of the wave. The filling factor is computed as the averaged ratio of the area occupied by the features and the total area of the corresponding FoV. It is worth noting that a non-unity filling factor can result in over- or under-estimation of the energy flux \citep{2008ApJ...676L..73V,2013ApJ...768..191G}. The densities are taken directly from the simulation as the average of the mean values within circles of 3 times the radii of each feature excluding the densities inside the feature, being $\approx 2.5\times 10^{-8}$\,kg\,m$^{-3}$ for Band 3 and $\approx 4.0\times 10^{-7}$\,kg\,m$^{-3}$ for Band~6. The factor 1/2 was introduced to account for the time-averaged energy over one period. For this study, the estimated energy fluxes correspond to those carried by the sausage modes. 

The radial velocity, $\mathrm{v_r}$, has previously been derived to estimate the properties of MHD sausage waves observed in pores \citep{2016ApJ...817...44F}. It is derived from the radial change of the feature under inspection ($\delta r$) and the oscillation period of the feature size ($P$). For the energy flux estimate presented here, we set $\delta r$ to the averaged amplitude of the radius oscillation across all features and $P$ to the averaged mean periods of size oscillations across all features. This procedure is applied to both observations and simulations separately. The radial velocity can then be written as 

\begin{equation}
    \mathrm{v_r} = \dfrac{2\pi\delta r}{P} \ ,
    \label{equ:vel_rad}
\end{equation}

\noindent the value of $\mathrm{v_r}$ corresponds to the maximum amplitude and the time averaged value to be used in the equation~\eqref{equ:energy_flux} is given by $\langle \mathrm{v_r} \rangle = \mathrm{v_r}/\sqrt{2}$ \citep{2012NatCo...3.1315M}. The phase speed of the wave $\mathrm{v_{ph}}$ can be expressed as 

\begin{equation}
    \mathrm{v_{ph}} = \dfrac{\omega}{\kappa} = c_s \sqrt{\dfrac{\pm A - 1}{\pm A - 1 + (\gamma - 1)(h\nu/k_B T)}} \ , 
    \label{equ:phase_speed}
\end{equation}

\noindent where $\gamma$ is the ratio of specific heats, $h$ is the Planck constant, $\nu$~is the frequency of observation ($100$ \,GHz for Band 3 and $240$ \,GHz for Band 6), $k_B$ is the Boltzmann constant, $T$ is the temperature within the feature and $c_s$ is the sound speed estimated using the average temperature of the observation or simulation in each case.
The dimensionless amplitude ratio  $A$ is defined as 

\begin{equation}
    A = \dfrac{\delta T/T_0}{\delta S/S_0} \ ,
    \label{equ:Temp_Area}
\end{equation}

\noindent where $\delta T$ is the amplitude of the temperature perturbation, $T_0$ is the mean temperature, $\delta S$ is the amplitude of the cross sectional area perturbation and $S_0$ is the mean area. \citet{2015ApJ...806..132G} defined and used Eq.~\eqref{equ:phase_speed} to estimate the phase speed of upwardly propagating oscillations in a pore at specific periods observed in distinct wavelengths. In their formalism, fractional changes in emergent intensity from ROSA images are used. Given that brightness temperature measured from ALMA lies within the Rayleigh-Jeans limit, it can be applied as a proxy for emission intensity.

Table~\ref{tab:computed_par} summarises the variables involved in the estimation of flux energy. Each value is shown with its respective uncertainty from the error propagation analysis. The large area perturbations in comparison to previous works \citep{2015ApJ...806..132G} tentatively suggest again that the dominant observed waves correspond to fast-sausage modes \citep{1983SoPh...88..179E}. The flux of energy in the Band 3 observation is{ 1839 $\pm$ 464\,W\,m$^{-2}$} while the energy flux in the Band 6 observation is {5341 $\pm$ 1348\,W\,m$^{-2}$}. For the {degraded} simulation, the energy fluxes are {621$\pm$ 407\,W\,m$^{-2}$} and {3640 $\pm$ 1557\,W\,m$^{-2}$} for bands 3 and 6 respectively. {For the original simulation, the energy fluxes are {453$\pm$ 289\,W\,m$^{-2}$} and {5485 $\pm$ 1587\,W\,m$^{-2}$} for bands 3 and 6 respectively.} In {all the} cases the energy flux decreases from Band 6 to Band 3. Using the median formation heights {of the degraded simulation} in Table~\ref{tab:basic_qtt_v2} of {1563\,km} for Band 3 and {1260\,km} for Band 6, the slope of the energy fluxes between the two heights {in the observation }is {12 $\pm$ 5\,Wm$^{-2}$ km$^{-1}$}.


\section{Discussion and Conclusions}
\label{sec:discussion} 
\label{sec:conclusions}

We have analysed {193} and {293} small bright features in ALMA observations{, 36 }and {392} features in corresponding synthetic continuum maps from Bifrost in ALMA bands 3 and 6 {at original resolution, and 24 for Band 3 and 204 for Band 6 at the degraded resolution}. It should be noted that the spatial resolution is critical for this type of study, and that any further loss of resolution would have considerable impact on the quality of data and the resulting detection of such features. The method used for the study was designed to ensure reliable detection of small-scale events down to sizes similar to the size of the ALMA beam. However, the method systematically underestimates the true size of the detected features by $\sim$20\% for most common sizes in Band 3 and 6. We note that it does not affect the derived oscillation periods, or the measured temperatures and {horizontal} velocities. {However, it might affect the radial velocity in Eq.~\eqref{equ:vel_rad} and therefore the estimated energy flux by a factor of $\sqrt{2}$.} The reliability of the method is discussed in more detail in Appendix~\ref{apx:A}. The fact that the method performed equally well for the observations and simulations tested in the study clearly demonstrates the overall reliability of this approach. 

The direct observation of MHD modes in the detected features implies that they are magnetic in nature. This interpretation is supported by the magnetic-field topology of the datasets analysed here, as demonstrated in \citet{2021RSPTA.37900174J}, indicating that the bright features are the result of magnetic fields branching from the solar surface into the chromosphere. Additional support is provided by the fact that the simulations exhibit very similar features which are clearly connected to magnetic field structures (i.e., they are cross sections of magnetic flux tubes). The properties of the features, namely, temperature, intensity and horizontal velocity change with time (see Fig.~\ref{fig:featexamp_snaps}) and exhibit an oscillatory behaviour (see Fig.~\ref{fig:featexam_cwt}). The high temporal resolution and unprecedented spatial resolution (at these wavelengths) of the ALMA observations allow reliable characterisation of the observed oscillations. It is found that the number of features per square mega-metre per second in the observations and simulations falls in the same order of magnitude for each corresponding receiver band, respectively (see Tab.~\ref{tab:basic_qtt_v2}). Interestingly, there are between 5-6 times mores features in Band~6 than in Band~3 in both observation and simulation. This may be due to several factors: (i)~Features in Band 3 with sizes comparable to the majority of features detected in Band 6 would not be observable because of the lower spatial resolution of Band 3. (ii)~On average, the continuum radiation in Band 6 is formed below the continuum in Band~3, while magnetic features tend to expand with height, directly affecting detection at a given resolution. (iii)~As a result of different formation heights, ALMA Band 3 and 6 are mapping different layers and thus different parts of the magnetic structures that are connected to the detected features. Correspondingly, different properties of the mapped magnetic environment would affect the occurrence of small-scale bright features.

The periods of oscillations in temperature, size, and velocity are identified by means of wavelet analysis (see Sec.~\ref{sec:analysis} and Fig.~\ref{fig:featexam_cwt}) and summarised in Table~\ref{tab:period}. The horizontal velocity amplitudes are summarised in Table~\ref{tab:basic_qtt_v2}. The determined values of velocity amplitudes and oscillation periods are comparable between those obtained from the simulations and observations in each respective band. On top of that, the calculated parameters fall within typical values of MHD modes observed in distinct features, that is respectively, 5-29~km\,s$^{-1}$ and 37-350~s for spicules \citet{2007Sci...318.1574D,2009ApJ...705L.217H,2012ApJ...744L...5J,2012ApJ...759...18P,2019A&A...631L...5B}, 1-10~km\,s$^{-1}$ and 100-250~s \citet{2013A&A...549A.116J,2013A&A...554A.115S} for bright points, 8-11~kms$^{-1}$ and 120-180~s for mottles \cite{2012ApJ...750...51K,2013ApJ...779...82K} and for fibrils 1-7~km\,s$^{-1}$ and 94-315~s \citet{2011ApJ...739...92P,2012NatCo...3.1315M}. Although it is not possible to unambiguously determine what type of the aforementioned phenomena we are observing in these ALMA datasets, the agreement of the derived oscillation periods and velocity amplitudes supports the conclusion that the detected oscillations are indeed associated to MHD modes. Furthermore, we speculate that the oscillations in horizontal velocities may be associated to MHD transverse (kink) waves.

{Moreover, oscillation periods larger than 100\,s (corresponding to frequencies lower than 10\,mHz) seem to be absent in Band~6 but present in Band~3 (see Fig.~\ref{fig:periods}) with the mean and median periods being accordingly shorter in Band~6 (see Table~\ref{tab:period}). This finding is consistent with other studies using other observing facilities at different wavelengths and different spatial resolutions. For instance, high-frequency waves (> 10\,mHz) in the lower-to-mid chromosphere are observed in slender fibrillar structures \citep[see e.g.][]{2017ApJS..229....7G,2017ApJS..229....9J}, whereas lower frequencies (< 10 mHz) are often reported to be present in the high chromosphere, e.g., for fibrillar structures observed in the H$\alpha$  and Ca II 854.2\,nm lines \citep[see e.g.,][]{2007ApJ...655..624D,2012ApJ...750...51K,2012A&A...543A...6M}. However, we should note that no direct observational evidence for the dissipation of energy associated with high-frequency waves has been found to date. While the estimated decrease of the energy flux with height of the identified sausage waves from Band 6 to Band 3 could suggest dissipation of wave energy between the respective formation height ranges of the two bands, the fact that the ALMA data for Band~3 and Band~6 used here were not obtained simultaneous prevents drawing any strong conclusion regarding the differences in the found oscillations period distributions and its implications for any related decrease in the energy flux.}

The analysis of phase differences between temperature and size oscillations shows a predominant anti-phase behaviour (see Fig.~\ref{fig:phases}). The phase differences together with the variation of the cross-sectional area of the features could indicate the presence of compressible MHD fast-sausage modes  \citep{2013A&A...555A..75M,2012NatCo...3.1315M,2021RSPTA.37900184G},  while the in-phase relation may be an indication of slow-sausage modes. In the real Sun, there is a mix of superimposed modes that are challenging to disentangle when observed, but whose particular signatures have nevertheless been found in the ALMA observations. {Which is reflected as phases out of the anti-phase and in-phase behaviours. From the simulations, we have found that these phases are} mostly coming from the regions with low magnetic-field strength, i.e., towards the fainter parts in Fig.~\ref{fig:sim_convbeam}. Thus, future works should consider analyses of other physical quantities that the simulations provide and the observations lack (such as gas density, pressure, and magnetic fields), to further explore the properties of MHD modes. Additionally, simulations can provide further information on how spatial resolution can influence the detection of such wave modes co-existing across the same small-scale magnetic structures.

In addition, we present the first estimation of the energy flux of sausage modes in numerous structures at chromospheric heights using ALMA observations. For this, we have made use of Eq.~\eqref{equ:energy_flux} in which all quantities are derived directly from our measurements except for the mass density. Hence, we have used the densities directly extracted from the simulations as explained in Sect.~\ref{subsec:wave_energy}. As expected, the energy fluxes in Band 6, i.e., lower in the atmosphere, are larger than the energy fluxes in Band 3, i.e., higher in the atmosphere. This implies that, on average, the energy transported upwards through the chromosphere (along the small magnetic structures) is damped at these heights, thus suggesting the propagation of these waves in the solar chromosphere. {The propagation of MHD waves between different layers corresponding to two frequencies within the Band 3 observations has been proven by \cite{2022A&A...665L...2G} who showed that transverse waves propagate with speeds of the order of 96\,km\,s$^{-1}$ and carry a flux of energy of the order of 4000\,W\,m$^{-2}$. Although the propagation of waves cannot be measured directly, the observations presented here imply an estimated energy flux of the fast-sausage mode of  1838$\pm$429\,W\,m$^{-2}$ and 5341$\pm$1448\,W\,m$^{-2}$} in bands 3 and 6, respectively. The energy fluxes from the {degraded} simulations were approximated as {621$\pm$407\,W\,m$^{-2}$ in Band 3 and 3640$\pm$1557\,W\,m$^{-2}$ in Band 6. For the simulation at the original resolutions the energy fluxes were estimated as 453$\pm$289\,W\,m$^{-2}$ in Band 3 and 5485$\pm$1587\,W\,m$^{-2}$ in Band 6.} The energy fluxes computed from both observations and simulations in Band~6 are favourably comparable within their uncertainties. The relatively higher difference of energies between the observations and degraded simulations in Band~3 may be due to the lower spatial resolution which consequently has an impact on the measured quantities, but this may only result in relatively larger uncertainties. {In fact, the energy flux in Band 6 at the original resolution is just 3\% bigger than the energy flux in the Band 6 observations.}

{{The structures presented in this paper, exhibit lower energies to wave modes detected in similar small chromospheric flux tubes. Alfv{\'{e}}n waves exhibiting energy flux of ~15~\,kW\,m$^{-2}$ \citep{2009Sci...323.1582J} and kink modes with energies in excess of 30000\,W\,m$^{-2}$ \citep{2022ApJ...930..129B} have been detected in fibrils and spicules across the chromosphere. In particular, \citet{2012NatCo...3.1315M} estimated the wave energy flux of compressible MHD sausage waves, presented in 103 chromospheric structures observed in the H$\alpha$ line-core, as 11700$\pm$3800\,W\,m$^{-2}$. With the H$\alpha$ spectral line forming in the mid-to-upper chromosphere, close to the heights sampled by ALMA \citep{2017A&A...598A..89R,2019ApJ...881...99M,2021A&A...651A...6B,2021ApJ...920..125M}, this infers that the detected waves are less energetic than their H$\alpha$ counterparts. However, a robust comparison of ALMA band formation heights compared to other chromospheric diagnostics is lacking \citep{2019ApJ...881...99M} so the energy differential may be influenced by formation height offsets.
Moreover, it is worth noting that kink modes with relatively low energies on the order of 310\,W\,m$^{-2}$ detected in small-scale magnetic elements \citep{2013A&A...549A.116J} as well as 15\,kW\,m$^{-2}$ in slender fibrillar structures \citep{2017ApJS..229....9J} have also been reported for observations in the low-to-mid chromosphere, sampled by the {\sc Sunrise} Ca\,{\sc ii}\,H filtergrams \citep{2010ApJ...723L.127S,2017ApJS..229....2S}.

In light of estimated chromospheric radiative losses of between 1000 - 10,000\,W\,m$^{-2}$ \citep{Withbroe1977}, it is unlikely that these waves alone can provide sufficient energy to balance the chromosphere. However, as part of a plethora of energetic chromospheric waves modes, they may play a part. Indeed, a difference in energy between the waves when observed in ALMA bands 3 and 6 could be a signature of damping (see Sect.~\ref{subsec:wave_energy}). Although, with the observations not being co-temporal, and with no propagation statistics, it cannot be concluded whether this is damping in identical wave trains due to upward propagation.  It is still a reasonable possibility that this wave energy is damped in the chromosphere, as significant damping has already been observed in sausage modes bridging into the chromosphere \citep{2015ApJ...806..132G, 2021RSPTA.37900172G}. In terms of chromospheric heating, any damped energy would need to be dissipated into localised plasma heating for these waves to have an influence. The obvious candidate would be leaky modes which naturally deposit energy \citep{1986SoPh..103..277C}, as fast modes can readily take this form, though the fast mode classification of waves in this study is only tentative. Apart from this, compressible waves have been proposed to dissipate their energies by multiple mechanisms, notably mode conversion \citep{1991A&A...241..625U}, resonant absorption \citep{doi:10.1063/1.1343090}, thermal conduction  \citep{2003SPD....34.0108O} and turbulent mixing \citep{2011ApJ...736....3V}. These mechanisms act on scales below current resolution capabilities, so remain a viable, but hypothetical, method for direct dissipation. There is incentive therefore for future detailed studies of wave propagation in ALMA observations. In particular, wave mode characterisation, alongside propagation between sub-bands  would be invaluable to accurately derive wave energetics and their influence on plasma heating, given the large population of these small flux tubes across the solar atmosphere.}}

Further research on how to exploit ALMA's capability of measuring  temperatures, as well as the use of diagnostics at other wavelengths is necessary to unveil the potential role of the energy carried by these waves in the heating. Furthermore, the possibility of co-simultaneous observations of the same target at two different ALMA bands is also a direct desired conclusion of this study, as it would facilitate the study of the same feature at different atmospheric heights, resulting in more direct conclusions regarding the energy transport and heating in such small-scale structures. 

\section*{Acknowledgments}
J.C.G.G wish to thank Carlos José Díaz Baso for his valuable input on the ML clustering method. This work is supported by the SolarALMA project, which received funding from the European Research Council (ERC) under the European Union’s Horizon 2020 research and innovation programme (grant agreement No. 682462), and by the Research Council of Norway through its Centres of Excellence scheme, project number 262622. HE was supported through the CHROMATIC project (2016.0019) funded by the Knut and Alice Wallenberg foundation.
This paper makes use of the following ALMA data: ADS/JAO.ALMA\#2016.1.00050.S. ALMA is a partnership of ESO (representing its member states), NSF (USA) and NINS (Japan), together with NRC(Canada), MOST and ASIAA (Taiwan), and KASI (Republic of Korea), in co-operation with the Republic of Chile. The Joint ALMA Observatory is operated by ESO, AUI/NRAO and NAOJ. We are grateful to the many colleagues who contributed to developing the solar observing modes for ALMA and for support from the ALMA Regional Centres. S.D.T.G. is grateful to the UK STFC for the consolidated grant ST/T00021X/1. Collaboration between authors was facilitated by support from the International Space Science Institute, Bern within the International Teams program (Team 502 -- WaLSA: Waves in the Lower Solar Atmosphere). Scientific discussions with the WaLSA team (\href{https://www.WaLSA.team}{https://www.WaLSA.team}) are acknowledged.


\bibliographystyle{aa}
\bibliography{ref.bib}

\begin{thebibliography}{95}
\expandafter\ifx\csname natexlab\endcsname\relax\def\natexlab#1{#1}\fi

\bibitem[{{Bate} {et~al.}(2022){Bate}, {Jess}, {Nakariakov}, {Grant},
  {Jafarzadeh}, {Stangalini}, {Keys}, {Christian}, \&
  {Keenan}}]{2022ApJ...930..129B}
{Bate}, W., {Jess}, D.~B., {Nakariakov}, V.~M., {et~al.} 2022, \apj, 930, 129

\bibitem[{{Bloomfield} {et~al.}(2004){Bloomfield}, {McAteer}, {Mathioudakis},
  {Williams}, \& {Keenan}}]{2004ApJ...604..936B}
{Bloomfield}, D.~S., {McAteer}, R.~T.~J., {Mathioudakis}, M., {Williams},
  D.~R., \& {Keenan}, F.~P. 2004, \apj, 604, 936

\bibitem[{{Bose} {et~al.}(2019){Bose}, {Henriques}, {Joshi}, \& {Rouppe van der
  Voort}}]{2019A&A...631L...5B}
{Bose}, S., {Henriques}, V. M.~J., {Joshi}, J., \& {Rouppe van der Voort}, L.
  2019, \aap, 631, L5

\bibitem[{{Braj{\v{s}}a} {et~al.}(2021){Braj{\v{s}}a}, {Skoki{\'c}}, {Sudar},
  {Benz}, {Krucker}, {Ludwig}, {Saar}, \& {Selhorst}}]{2021A&A...651A...6B}
{Braj{\v{s}}a}, R., {Skoki{\'c}}, I., {Sudar}, D., {et~al.} 2021, \aap, 651, A6

\bibitem[{{Cally}(1986)}]{1986SoPh..103..277C}
{Cally}, P.~S. 1986, \solphys, 103, 277

\bibitem[{{Carlsson} {et~al.}(2016){Carlsson}, {Hansteen}, {Gudiksen},
  {Leenaarts}, \& {De Pontieu}}]{2016A&A...585A...4C}
{Carlsson}, M., {Hansteen}, V.~H., {Gudiksen}, B.~V., {Leenaarts}, J., \& {De
  Pontieu}, B. 2016, \aap, 585, A4

\bibitem[{{Crocker} \& {Grier}(1996)}]{1996JCIS..179..298C}
{Crocker}, J.~C. \& {Grier}, D.~G. 1996, Journal of Colloid and Interface
  Science, 179, 298

\bibitem[{de~la Cruz~Rodríguez {et~al.}(2021)de~la Cruz~Rodríguez,
  Szydlarski, \& Wedemeyer}]{2021_art}
de~la Cruz~Rodríguez, J., Szydlarski, M., \& Wedemeyer, S. 2021, {ART:
  Advanced (and fast!) Radiative Transfer code for Solar Physics
  (https://github.com/SolarAlma/ART).}

\bibitem[{{de Moortel}(2009)}]{2009SSRv..149...65D}
{de Moortel}, I. 2009, \ssr, 149, 65

\bibitem[{{De Pontieu} {et~al.}(2007{\natexlab{a}}){De Pontieu}, {Hansteen},
  {Rouppe van der Voort}, {van Noort}, \& {Carlsson}}]{2007ApJ...655..624D}
{De Pontieu}, B., {Hansteen}, V.~H., {Rouppe van der Voort}, L., {van Noort},
  M., \& {Carlsson}, M. 2007{\natexlab{a}}, \apj, 655, 624

\bibitem[{{De Pontieu} {et~al.}(2007{\natexlab{b}}){De Pontieu}, {McIntosh},
  {Carlsson}, {Hansteen}, {Tarbell}, {Schrijver}, {Title}, {Shine}, {Tsuneta},
  {Katsukawa}, {Ichimoto}, {Suematsu}, {Shimizu}, \&
  {Nagata}}]{2007Sci...318.1574D}
{De Pontieu}, B., {McIntosh}, S.~W., {Carlsson}, M., {et~al.}
  2007{\natexlab{b}}, Science, 318, 1574

\bibitem[{{Dorotovi{\v{c}}} {et~al.}(2014){Dorotovi{\v{c}}}, {Erd{\'e}lyi},
  {Freij}, {Karlovsk{\'y}}, \& {M{\'a}rquez}}]{2014A&A...563A..12D}
{Dorotovi{\v{c}}}, I., {Erd{\'e}lyi}, R., {Freij}, N., {Karlovsk{\'y}}, V., \&
  {M{\'a}rquez}, I. 2014, \aap, 563, A12

\bibitem[{{Dulk}(1985)}]{1985ARA&A..23..169D}
{Dulk}, G.~A. 1985, \araa, 23, 169

\bibitem[{{Edwin} \& {Roberts}(1983)}]{1983SoPh...88..179E}
{Edwin}, P.~M. \& {Roberts}, B. 1983, \solphys, 88, 179

\bibitem[{{Eklund} {et~al.}(2021{\natexlab{a}}){Eklund}, {Wedemeyer}, {Snow},
  {Jess}, {Jafarzadeh}, {Grant}, {Carlsson}, \&
  {Szydlarski}}]{2021RSPTA.37900185E}
{Eklund}, H., {Wedemeyer}, S., {Snow}, B., {et~al.} 2021{\natexlab{a}},
  Philosophical Transactions of the Royal Society of London Series A, 379,
  20200185

\bibitem[{{Eklund} {et~al.}(2021{\natexlab{b}}){Eklund}, {Wedemeyer},
  {Szydlarski}, \& {Jafarzadeh}}]{2021A&A...656A..68E}
{Eklund}, H., {Wedemeyer}, S., {Szydlarski}, M., \& {Jafarzadeh}, S.
  2021{\natexlab{b}}, \aap, 656, A68

\bibitem[{{Eklund} {et~al.}(2020){Eklund}, {Wedemeyer}, {Szydlarski},
  {Jafarzadeh}, \& {Guevara G{\'o}mez}}]{2020A&A...644A.152E}
{Eklund}, H., {Wedemeyer}, S., {Szydlarski}, M., {Jafarzadeh}, S., \& {Guevara
  G{\'o}mez}, J.~C. 2020, \aap, 644, A152

\bibitem[{{Freij} {et~al.}(2016){Freij}, {Dorotovi{\v{c}}}, {Morton},
  {Ruderman}, {Karlovsk{\'y}}, \& {Erd{\'e}lyi}}]{2016ApJ...817...44F}
{Freij}, N., {Dorotovi{\v{c}}}, I., {Morton}, R.~J., {et~al.} 2016, \apj, 817,
  44

\bibitem[{{Gafeira} {et~al.}(2017){Gafeira}, {Jafarzadeh}, {Solanki}, {Lagg},
  {van Noort}, {Barthol}, {Blanco Rodr{\'\i}guez}, {del Toro Iniesta},
  {Gandorfer}, {Gizon}, {Hirzberger}, {Kn{\"o}lker}, {Orozco Su{\'a}rez},
  {Riethm{\"u}ller}, \& {Schmidt}}]{2017ApJS..229....7G}
{Gafeira}, R., {Jafarzadeh}, S., {Solanki}, S.~K., {et~al.} 2017, \apjs, 229, 7

\bibitem[{{Gilchrist-Millar} {et~al.}(2021){Gilchrist-Millar}, {Jess}, {Grant},
  {Keys}, {Beck}, {Jafarzadeh}, {Riedl}, {Van Doorsselaere}, \& {Ruiz
  Cobo}}]{2021RSPTA.37900172G}
{Gilchrist-Millar}, C.~A., {Jess}, D.~B., {Grant}, S. D.~T., {et~al.} 2021,
  \ptrsa, 379, 20200172

\bibitem[{Goossens \& De~Groof(2001)}]{doi:10.1063/1.1343090}
Goossens, M. \& De~Groof, A. 2001, Physics of Plasmas, 8, 2371

\bibitem[{{Goossens} {et~al.}(2013){Goossens}, {Van Doorsselaere}, {Soler}, \&
  {Verth}}]{2013ApJ...768..191G}
{Goossens}, M., {Van Doorsselaere}, T., {Soler}, R., \& {Verth}, G. 2013, \apj,
  768, 191

\bibitem[{{Grant} {et~al.}(2015){Grant}, {Jess}, {Moreels}, {Morton},
  {Christian}, {Giagkiozis}, {Verth}, {Fedun}, {Keys}, {Van Doorsselaere}, \&
  {Erd{\'e}lyi}}]{2015ApJ...806..132G}
{Grant}, S.~D.~T., {Jess}, D.~B., {Moreels}, M.~G., {et~al.} 2015, \apj, 806,
  132

\bibitem[{{Grant} {et~al.}(2022){Grant}, {Jess}, {Stangalini}, {Jafarzadeh},
  {Fedun}, {Verth}, {Keys}, {Rajaguru}, {Uitenbroek}, {MacBride}, {Bate}, \&
  {Gilchrist-Millar}}]{2022ApJ...938..143G}
{Grant}, S.~D.~T., {Jess}, D.~B., {Stangalini}, M., {et~al.} 2022, \apj, 938,
  143

\bibitem[{{Grant} {et~al.}(2018){Grant}, {Jess}, {Zaqarashvili}, {Beck},
  {Socas-Navarro}, {Aschwanden}, {Keys}, {Christian}, {Houston}, \&
  {Hewitt}}]{2018NatPh..14..480G}
{Grant}, S. D.~T., {Jess}, D.~B., {Zaqarashvili}, T.~V., {et~al.} 2018, Nature
  Physics, 14, 480

\bibitem[{{Gudiksen} {et~al.}(2011){Gudiksen}, {Carlsson}, {Hansteen}, {Hayek},
  {Leenaarts}, \& {Mart{\'\i}nez-Sykora}}]{2011A&A...531A.154G}
{Gudiksen}, B.~V., {Carlsson}, M., {Hansteen}, V.~H., {et~al.} 2011, \aap, 531,
  A154

\bibitem[{{Guevara G{\'o}mez} {et~al.}(2022){Guevara G{\'o}mez}, {Jafarzadeh},
  {Wedemeyer}, \& {Szydlarski}}]{2022A&A...665L...2G}
{Guevara G{\'o}mez}, J.~C., {Jafarzadeh}, S., {Wedemeyer}, S., \& {Szydlarski},
  M. 2022, \aap, 665, L2

\bibitem[{{Guevara G{\'o}mez} {et~al.}(2021){Guevara G{\'o}mez}, {Jafarzadeh},
  {Wedemeyer}, {Szydlarski}, {Stangalini}, {Fleck}, \&
  {Keys}}]{2021RSPTA.37900184G}
{Guevara G{\'o}mez}, J.~C., {Jafarzadeh}, S., {Wedemeyer}, S., {et~al.} 2021,
  Philosophical Transactions of the Royal Society of London Series A, 379,
  20200184

\bibitem[{{He} {et~al.}(2009){He}, {Marsch}, {Tu}, \&
  {Tian}}]{2009ApJ...705L.217H}
{He}, J., {Marsch}, E., {Tu}, C., \& {Tian}, H. 2009, \apjl, 705, L217

\bibitem[{{Henriques} {et~al.}(2022){Henriques}, {Jafarzadeh}, {Guevara
  G{\'o}mez}, {Eklund}, {Wedemeyer}, {Szydlarski}, {Haugan}, \&
  {Mohan}}]{2022A&A...659A..31H}
{Henriques}, V. M.~J., {Jafarzadeh}, S., {Guevara G{\'o}mez}, J.~C., {et~al.}
  2022, \aap, 659, A31

\bibitem[{{Ionson}(1978)}]{1978ApJ...226..650I}
{Ionson}, J.~A. 1978, \apj, 226, 650

\bibitem[{{Jafarzadeh} {et~al.}(2021{\natexlab{a}}){Jafarzadeh}, {Guevara
  Gómez}, {Wedemeyer}, \& {Aannerud}}]{shahin_jafarzadeh_2021_5466873}
{Jafarzadeh}, S., {Guevara Gómez}, J.~C., {Wedemeyer}, S., \& {Aannerud}, S.
  2021{\natexlab{a}}, SolarAlma/SALAT: v1.0

\bibitem[{{Jafarzadeh} {et~al.}(2013){Jafarzadeh}, {Solanki}, {Feller}, {Lagg},
  {Pietarila}, {Danilovic}, {Riethm{\"u}ller}, \& {Mart{\'\i}nez
  Pillet}}]{2013A&A...549A.116J}
{Jafarzadeh}, S., {Solanki}, S.~K., {Feller}, A., {et~al.} 2013, \aap, 549,
  A116

\bibitem[{{Jafarzadeh} {et~al.}(2017{\natexlab{a}}){Jafarzadeh}, {Solanki},
  {Gafeira}, {van Noort}, {Barthol}, {Blanco Rodr{\'\i}guez}, {del Toro
  Iniesta}, {Gandorfer}, {Gizon}, {Hirzberger}, {Kn{\"o}lker}, {Orozco
  Su{\'a}rez}, {Riethm{\"u}ller}, \& {Schmidt}}]{2017ApJS..229....9J}
{Jafarzadeh}, S., {Solanki}, S.~K., {Gafeira}, R., {et~al.} 2017{\natexlab{a}},
  \apjs, 229, 9

\bibitem[{{Jafarzadeh} {et~al.}(2017{\natexlab{b}}){Jafarzadeh}, {Solanki},
  {Stangalini}, {Steiner}, {Cameron}, \& {Danilovic}}]{2017ApJS..229...10J}
{Jafarzadeh}, S., {Solanki}, S.~K., {Stangalini}, M., {et~al.}
  2017{\natexlab{b}}, \apjs, 229, 10

\bibitem[{{Jafarzadeh} {et~al.}(2021{\natexlab{b}}){Jafarzadeh}, {Wedemeyer},
  {Fleck}, {Stangalini}, {Jess}, {Morton}, {Szydlarski}, {Henriques}, {Zhu},
  {Wiegelmann}, {Guevara G{\'o}mez}, {Grant}, {Chen}, {Reardon}, \&
  {White}}]{2021RSPTA.37900174J}
{Jafarzadeh}, S., {Wedemeyer}, S., {Fleck}, B., {et~al.} 2021{\natexlab{b}},
  Philosophical Transactions of the Royal Society of London Series A, 379,
  20200174

\bibitem[{{Jess} {et~al.}(2007){Jess}, {Andi{\'c}}, {Mathioudakis},
  {Bloomfield}, \& {Keenan}}]{2007A&A...473..943J}
{Jess}, D.~B., {Andi{\'c}}, A., {Mathioudakis}, M., {Bloomfield}, D.~S., \&
  {Keenan}, F.~P. 2007, \aap, 473, 943

\bibitem[{{Jess} {et~al.}(2022){Jess}, {Jafarzadeh}, {Keys}, {Stangalini},
  {Verth}, \& {Grant}}]{2022LRSP...00..000J}
{Jess}, D.~B., {Jafarzadeh}, S., {Keys}, P.~H., {et~al.} 2022, \lrsp, in press
  [\eprint[arXiv]{2212.09788}]

\bibitem[{{Jess} {et~al.}(2010){Jess}, {Mathioudakis}, {Christian}, {Keenan},
  {Ryans}, \& {Crockett}}]{2010SoPh..261..363J}
{Jess}, D.~B., {Mathioudakis}, M., {Christian}, D.~J., {et~al.} 2010, \solphys,
  261, 363

\bibitem[{{Jess} {et~al.}(2009){Jess}, {Mathioudakis}, {Erd{\'e}lyi},
  {Crockett}, {Keenan}, \& {Christian}}]{2009Sci...323.1582J}
{Jess}, D.~B., {Mathioudakis}, M., {Erd{\'e}lyi}, R., {et~al.} 2009, Science,
  323, 1582

\bibitem[{{Jess} {et~al.}(2015){Jess}, {Morton}, {Verth}, {Fedun}, {Grant}, \&
  {Giagkiozis}}]{2015SSRv..190..103J}
{Jess}, D.~B., {Morton}, R.~J., {Verth}, G., {et~al.} 2015, \ssr, 190, 103

\bibitem[{{Jess} {et~al.}(2012){Jess}, {Pascoe}, {Christian}, {Mathioudakis},
  {Keys}, \& {Keenan}}]{2012ApJ...744L...5J}
{Jess}, D.~B., {Pascoe}, D.~J., {Christian}, D.~J., {et~al.} 2012, \apjl, 744,
  L5

\bibitem[{{Jess} \& {Verth}(2016)}]{2016GMS...216..449J}
{Jess}, D.~B. \& {Verth}, G. 2016, Washington DC American Geophysical Union
  Geophysical Monograph Series, 216, 449

\bibitem[{{Keys} {et~al.}(2018){Keys}, {Morton}, {Jess}, {Verth}, {Grant},
  {Mathioudakis}, {Mackay}, {Doyle}, {Christian}, {Keenan}, \&
  {Erd{\'e}lyi}}]{2018ApJ...857...28K}
{Keys}, P.~H., {Morton}, R.~J., {Jess}, D.~B., {et~al.} 2018, \apj, 857, 28

\bibitem[{{Klimchuk}(2006)}]{2006SoPh..234...41K}
{Klimchuk}, J.~A. 2006, \solphys, 234, 41

\bibitem[{{Kohutova} \& {Popovas}(2021)}]{2021A&A...647A..81K}
{Kohutova}, P. \& {Popovas}, A. 2021, \aap, 647, A81

\bibitem[{{Kuridze} {et~al.}(2012){Kuridze}, {Morton}, {Erd{\'e}lyi},
  {Dorrian}, {Mathioudakis}, {Jess}, \& {Keenan}}]{2012ApJ...750...51K}
{Kuridze}, D., {Morton}, R.~J., {Erd{\'e}lyi}, R., {et~al.} 2012, \apj, 750, 51

\bibitem[{{Kuridze} {et~al.}(2013){Kuridze}, {Verth}, {Mathioudakis},
  {Erd{\'e}lyi}, {Jess}, {Morton}, {Christian}, \&
  {Keenan}}]{2013ApJ...779...82K}
{Kuridze}, D., {Verth}, G., {Mathioudakis}, M., {et~al.} 2013, \apj, 779, 82

\bibitem[{{Leenaarts} {et~al.}(2015){Leenaarts}, {Carlsson}, \& {Rouppe van der
  Voort}}]{2015ApJ...802..136L}
{Leenaarts}, J., {Carlsson}, M., \& {Rouppe van der Voort}, L. 2015, \apj, 802,
  136

\bibitem[{{Lemen} {et~al.}(2012){Lemen}, {Title}, {Akin}, {Boerner}, {Chou},
  {Drake}, {Duncan}, {Edwards}, {Friedlaender}, {Heyman}, {Hurlburt}, {Katz},
  {Kushner}, {Levay}, {Lindgren}, {Mathur}, {McFeaters}, {Mitchell}, {Rehse},
  {Schrijver}, {Springer}, {Stern}, {Tarbell}, {Wuelser}, {Wolfson}, {Yanari},
  {Bookbinder}, {Cheimets}, {Caldwell}, {Deluca}, {Gates}, {Golub}, {Park},
  {Podgorski}, {Bush}, {Scherrer}, {Gummin}, {Smith}, {Auker}, {Jerram}, \&
  et~al.}]{2012SoPh..275...17L}
{Lemen}, J.~R., {Title}, A.~M., {Akin}, D.~J., {et~al.} 2012, \solphys, 275, 17

\bibitem[{{Lin} {et~al.}(2007){Lin}, {Engvold}, {Rouppe van der Voort}, \& {van
  Noort}}]{2007SoPh..246...65L}
{Lin}, Y., {Engvold}, O., {Rouppe van der Voort}, L.~H.~M., \& {van Noort}, M.
  2007, \solphys, 246, 65

\bibitem[{{Loukitcheva} {et~al.}(2015){Loukitcheva}, {Solanki}, {Carlsson}, \&
  {White}}]{2015A&A...575A..15L}
{Loukitcheva}, M., {Solanki}, S.~K., {Carlsson}, M., \& {White}, S.~M. 2015,
  \aap, 575, A15

\bibitem[{{Molnar} {et~al.}(2019){Molnar}, {Reardon}, {Chai}, {Gary},
  {Uitenbroek}, {Cauzzi}, \& {Cranmer}}]{2019ApJ...881...99M}
{Molnar}, M.~E., {Reardon}, K.~P., {Chai}, Y., {et~al.} 2019, \apj, 881, 99

\bibitem[{{Molnar} {et~al.}(2021){Molnar}, {Reardon}, {Cranmer}, {Kowalski},
  {Chai}, \& {Gary}}]{2021ApJ...920..125M}
{Molnar}, M.~E., {Reardon}, K.~P., {Cranmer}, S.~R., {et~al.} 2021, \apj, 920,
  125

\bibitem[{{Moreels} {et~al.}(2013){Moreels}, {Goossens}, \& {Van
  Doorsselaere}}]{2013A&A...555A..75M}
{Moreels}, M.~G., {Goossens}, M., \& {Van Doorsselaere}, T. 2013, \aap, 555,
  A75

\bibitem[{{Moreels} {et~al.}(2015){Moreels}, {Van Doorsselaere}, {Grant},
  {Jess}, \& {Goossens}}]{2015A&A...578A..60M}
{Moreels}, M.~G., {Van Doorsselaere}, T., {Grant}, S.~D.~T., {Jess}, D.~B., \&
  {Goossens}, M. 2015, \aap, 578, A60

\bibitem[{{Morton}(2012)}]{2012A&A...543A...6M}
{Morton}, R.~J. 2012, \aap, 543, A6

\bibitem[{{Morton} {et~al.}(2011){Morton}, {Erd{\'e}lyi}, {Jess}, \&
  {Mathioudakis}}]{2011ApJ...729L..18M}
{Morton}, R.~J., {Erd{\'e}lyi}, R., {Jess}, D.~B., \& {Mathioudakis}, M. 2011,
  \apjl, 729, L18

\bibitem[{{Morton} {et~al.}(2012){Morton}, {Verth}, {Jess}, {Kuridze},
  {Ruderman}, {Mathioudakis}, \& {Erd{\'e}lyi}}]{2012NatCo...3.1315M}
{Morton}, R.~J., {Verth}, G., {Jess}, D.~B., {et~al.} 2012, Nature
  Communications, 3, 1315

\bibitem[{{Nakariakov} {et~al.}(2005){Nakariakov}, {Pascoe}, \&
  {Arber}}]{2005SSRv..121..115N}
{Nakariakov}, V.~M., {Pascoe}, D.~J., \& {Arber}, T.~D. 2005, \ssr, 121, 115

\bibitem[{{Narang} {et~al.}(2020){Narang}, {Chandrashekhar}, {Jafarzadeh},
  {Wedemeyer}, {Fleck}, \& {Szydlarski}}]{Narang2020}
{Narang}, N., {Chandrashekhar}, K., {Jafarzadeh}, S., {et~al.} 2020, \ptrsa

\bibitem[{{Nindos} {et~al.}(2021){Nindos}, {Patsourakos}, {Alissandrakis}, \&
  {Bastian}}]{2021A&A...652A..92N}
{Nindos}, A., {Patsourakos}, S., {Alissandrakis}, C.~E., \& {Bastian}, T.~S.
  2021, \aap, 652, A92

\bibitem[{{Ofman} \& {Terradas}(2003)}]{2003SPD....34.0108O}
{Ofman}, L. \& {Terradas}, J. 2003, in AAS/Solar Physics Division Meeting,
  Vol.~34, AAS/Solar Physics Division Meeting \#34, 01.08

\bibitem[{{Osterbrock}(1961)}]{1961ApJ...134..347O}
{Osterbrock}, D.~E. 1961, \apj, 134, 347

\bibitem[{{Patsourakos} {et~al.}(2020){Patsourakos}, {Alissandrakis}, {Nindos},
  \& {Bastian}}]{2020A&A...634A..86P}
{Patsourakos}, S., {Alissandrakis}, C.~E., {Nindos}, A., \& {Bastian}, T.~S.
  2020, \aap, 634, A86

\bibitem[{Pedregosa {et~al.}(2011)Pedregosa, Varoquaux, Gramfort, Michel,
  Thirion, Grisel, Blondel, Prettenhofer, Weiss, Dubourg, Vanderplas, Passos,
  Cournapeau, Brucher, Perrot, \& Duchesnay}]{scikit-learn}
Pedregosa, F., Varoquaux, G., Gramfort, A., {et~al.} 2011, Journal of Machine
  Learning Research, 12, 2825

\bibitem[{{Pereira} {et~al.}(2012){Pereira}, {De Pontieu}, \&
  {Carlsson}}]{2012ApJ...759...18P}
{Pereira}, T. M.~D., {De Pontieu}, B., \& {Carlsson}, M. 2012, \apj, 759, 18

\bibitem[{{Pesnell} {et~al.}(2012){Pesnell}, {Thompson}, \&
  {Chamberlin}}]{2012SoPh..275....3P}
{Pesnell}, W.~D., {Thompson}, B.~J., \& {Chamberlin}, P.~C. 2012, \solphys,
  275, 3

\bibitem[{{Pietarila} {et~al.}(2011){Pietarila}, {Aznar Cuadrado},
  {Hirzberger}, \& {Solanki}}]{2011ApJ...739...92P}
{Pietarila}, A., {Aznar Cuadrado}, R., {Hirzberger}, J., \& {Solanki}, S.~K.
  2011, \apj, 739, 92

\bibitem[{{Porter} {et~al.}(1994){Porter}, {Klimchuk}, \&
  {Sturrock}}]{1994ApJ...435..482P}
{Porter}, L.~J., {Klimchuk}, J.~A., \& {Sturrock}, P.~A. 1994, \apj, 435, 482

\bibitem[{{Rutten}(2017)}]{2017A&A...598A..89R}
{Rutten}, R.~J. 2017, \aap, 598, A89

\bibitem[{{Scherrer} {et~al.}(2012){Scherrer}, {Schou}, {Bush}, {Kosovichev},
  {Bogart}, {Hoeksema}, {Liu}, {Duvall}, {Zhao}, {Title}, {Schrijver},
  {Tarbell}, \& {Tomczyk}}]{2012SoPh..275..207S}
{Scherrer}, P.~H., {Schou}, J., {Bush}, R.~I., {et~al.} 2012, \solphys, 275,
  207

\bibitem[{{Solanki} {et~al.}(2010){Solanki}, {Barthol}, {Danilovic}, {Feller},
  {Gandorfer}, {Hirzberger}, {Riethm{\"u}ller}, {Sch{\"u}ssler}, {Bonet},
  {Mart{\'\i}nez Pillet}, {del Toro Iniesta}, {Domingo}, {Palacios},
  {Kn{\"o}lker}, {Bello Gonz{\'a}lez}, {Berkefeld}, {Franz}, {Schmidt}, \&
  {Title}}]{2010ApJ...723L.127S}
{Solanki}, S.~K., {Barthol}, P., {Danilovic}, S., {et~al.} 2010, \apjl, 723,
  L127

\bibitem[{{Solanki} {et~al.}(2017){Solanki}, {Riethm{\"u}ller}, {Barthol},
  {Danilovic}, {Deutsch}, {Doerr}, {Feller}, {Gandorfer}, {Germerott}, {Gizon},
  {Grauf}, {Heerlein}, {Hirzberger}, {Kolleck}, {Lagg}, {Meller}, {Tomasch},
  {van Noort}, {Blanco Rodr{\'\i}guez}, {Gasent Blesa}, {Balaguer Jim{\'e}nez},
  {Del Toro Iniesta}, {L{\'o}pez Jim{\'e}nez}, {Orozco Suarez}, {Berkefeld},
  {Halbgewachs}, {Schmidt}, {{\'A}lvarez-Herrero}, {Sabau-Graziati}, {P{\'e}rez
  Grand e}, {Mart{\'\i}nez Pillet}, {Card}, {Centeno}, {Kn{\"o}lker}, \&
  {Lecinski}}]{2017ApJS..229....2S}
{Solanki}, S.~K., {Riethm{\"u}ller}, T.~L., {Barthol}, P., {et~al.} 2017,
  \apjs, 229, 2

\bibitem[{{Srivastava} {et~al.}(2021){Srivastava}, {Ballester}, {Cally},
  {Carlsson}, {Goossens}, {Jess}, {Khomenko}, {Mathioudakis}, {Murawski}, \&
  {Zaqarashvili}}]{2021JGRA..12629097S}
{Srivastava}, A.~K., {Ballester}, J.~L., {Cally}, P.~S., {et~al.} 2021, Journal
  of Geophysical Research (Space Physics), 126, e029097

\bibitem[{{Stangalini} {et~al.}(2014){Stangalini}, {Consolini}, {Berrilli}, {De
  Michelis}, \& {Tozzi}}]{2014A&A...569A.102S}
{Stangalini}, M., {Consolini}, G., {Berrilli}, F., {De Michelis}, P., \&
  {Tozzi}, R. 2014, \aap, 569, A102

\bibitem[{{Stangalini} {et~al.}(2017){Stangalini}, {Giannattasio},
  {Erd{\'e}lyi}, {Jafarzadeh}, {Consolini}, {Criscuoli}, {Ermolli},
  {Guglielmino}, \& {Zuccarello}}]{2017ApJ...840...19S}
{Stangalini}, M., {Giannattasio}, F., {Erd{\'e}lyi}, R., {et~al.} 2017, \apj,
  840, 19

\bibitem[{{Stangalini} {et~al.}(2015){Stangalini}, {Giannattasio}, \&
  {Jafarzadeh}}]{2015A&A...577A..17S}
{Stangalini}, M., {Giannattasio}, F., \& {Jafarzadeh}, S. 2015, \aap, 577, A17

\bibitem[{{Stangalini} {et~al.}(2013){Stangalini}, {Solanki}, {Cameron}, \&
  {Mart{\'\i}nez Pillet}}]{2013A&A...554A.115S}
{Stangalini}, M., {Solanki}, S.~K., {Cameron}, R., \& {Mart{\'\i}nez Pillet},
  V. 2013, \aap, 554, A115

\bibitem[{{Stein} \& {Leibacher}(1974)}]{1974ARA&A..12..407S}
{Stein}, R.~F. \& {Leibacher}, J. 1974, \araa, 12, 407

\bibitem[{{Torrence} \& {Compo}(1998)}]{1998BAMS...79...61T}
{Torrence}, C. \& {Compo}, G.~P. 1998, Bulletin of the American Meteorological
  Society, 79, 61

\bibitem[{{Ulmschneider} {et~al.}(1991){Ulmschneider}, {Zaehringer}, \&
  {Musielak}}]{1991A&A...241..625U}
{Ulmschneider}, P., {Zaehringer}, K., \& {Musielak}, Z.~E. 1991, \aap, 241, 625

\bibitem[{{Valle Silva} {et~al.}(2021){Valle Silva}, {Gim{\'e}nez de Castro},
  {Selhorst}, {Raulin}, \& {Valio}}]{2021MNRAS.500.1964V}
{Valle Silva}, J.~F., {Gim{\'e}nez de Castro}, C.~G., {Selhorst}, C.~L.,
  {Raulin}, J.~P., \& {Valio}, A. 2021, \mnras, 500, 1964

\bibitem[{{van Ballegooijen} {et~al.}(2011){van Ballegooijen}, {Asgari-Targhi},
  {Cranmer}, \& {DeLuca}}]{2011ApJ...736....3V}
{van Ballegooijen}, A.~A., {Asgari-Targhi}, M., {Cranmer}, S.~R., \& {DeLuca},
  E.~E. 2011, \apj, 736, 3

\bibitem[{van~der Walt {et~al.}(2014)van~der Walt, {S}ch\"onberger,
  {Nunez-Iglesias}, {B}oulogne, {W}arner, {Y}ager, {G}ouillart, {Y}u, \& the
  scikit-image contributors}]{scikit-image}
van~der Walt, S., {S}ch\"onberger, J.~L., {Nunez-Iglesias}, J., {et~al.} 2014,
  PeerJ, 2, e453

\bibitem[{{Van Doorsselaere} {et~al.}(2008){Van Doorsselaere}, {Nakariakov}, \&
  {Verwichte}}]{2008ApJ...676L..73V}
{Van Doorsselaere}, T., {Nakariakov}, V.~M., \& {Verwichte}, E. 2008, \apjl,
  676, L73

\bibitem[{{van Doorsselaere} {et~al.}(2009){van Doorsselaere}, {Verwichte}, \&
  {Terradas}}]{2009SSRv..149..299V}
{van Doorsselaere}, T., {Verwichte}, E., \& {Terradas}, J. 2009, \ssr, 149, 299

\bibitem[{{Verth} \& {Jess}(2016)}]{2016GMS...216..431V}
{Verth}, G. \& {Jess}, D.~B. 2016, Washington DC American Geophysical Union
  Geophysical Monograph Series, 216, 431

\bibitem[{{Wedemeyer} {et~al.}(2016){Wedemeyer}, {Bastian}, {Braj{\v{s}}a},
  {Hudson}, {Fleishman}, {Loukitcheva}, {Fleck}, {Kontar}, {De Pontieu},
  {Yagoubov}, {Tiwari}, {Soler}, {Black}, {Antolin}, {Scullion}, {Gun{\'a}r},
  {Labrosse}, {Ludwig}, {Benz}, {White}, {Hauschildt}, {Doyle}, {Nakariakov},
  {Ayres}, {Heinzel}, {Karlicky}, {Van Doorsselaere}, {Gary}, {Alissandrakis},
  {Nindos}, {Solanki}, {Rouppe van der Voort}, {Shimojo}, {Kato},
  {Zaqarashvili}, {Perez}, {Selhorst}, \& {Barta}}]{2016SSRv..200....1W}
{Wedemeyer}, S., {Bastian}, T., {Braj{\v{s}}a}, R., {et~al.} 2016, \ssr, 200, 1

\bibitem[{{Wedemeyer} {et~al.}(2022){Wedemeyer}, {Fleishman}, {de la Cruz
  Rodr{\'\i}guez}, {Gun{\'a}r}, {da Silva Santos}, {Antolin}, {Guevara
  G{\'o}mez}, {Szydlarski}, \& {Eklund}}]{2022FrASS...9.7878W}
{Wedemeyer}, S., {Fleishman}, G., {de la Cruz Rodr{\'\i}guez}, J., {et~al.}
  2022, Frontiers in Astronomy and Space Sciences, 9, 967878

\bibitem[{{Wedemeyer} {et~al.}(2020){Wedemeyer}, {Szydlarski}, {Jafarzadeh},
  {Eklund}, {Guevara Gomez}, {Bastian}, {Fleck}, {de la Cruz Rodriguez},
  {Rodger}, \& {Carlsson}}]{2020A&A...635A..71W}
{Wedemeyer}, S., {Szydlarski}, M., {Jafarzadeh}, S., {et~al.} 2020, \aap, 635,
  A71

\bibitem[{{White} {et~al.}(2017){White}, {Iwai}, {Phillips}, {Hills}, {Hirota},
  {Yagoubov}, {Siringo}, {Shimojo}, {Bastian}, {Hales}, {Sawada}, {Asayama},
  {Sugimoto}, {Marson}, {Kawasaki}, {Muller}, {Nakazato}, {Sugimoto},
  {Braj{\v{s}}a}, {Skoki{\'c}}, {B{\'a}rta}, {Kim}, {Remijan}, {de Gregorio},
  {Corder}, {Hudson}, {Loukitcheva}, {Chen}, {De Pontieu}, {Fleishmann},
  {Gary}, {Kobelski}, {Wedemeyer}, \& {Yan}}]{2017SoPh..292...88W}
{White}, S.~M., {Iwai}, K., {Phillips}, N.~M., {et~al.} 2017, \solphys, 292, 88

\bibitem[{{White} {et~al.}(2006){White}, {Loukitcheva}, \&
  {Solanki}}]{2006A&A...456..697W}
{White}, S.~M., {Loukitcheva}, M., \& {Solanki}, S.~K. 2006, \aap, 456, 697

\bibitem[{{Withbroe} \& {Noyes}(1977)}]{Withbroe1977}
{Withbroe}, G.~L. \& {Noyes}, R.~W. 1977, \araa, 15, 363

\bibitem[{{Wootten} \& {Thompson}(2009)}]{2009IEEEP..97.1463W}
{Wootten}, A. \& {Thompson}, A.~R. 2009, IEEE Proceedings, 97, 1463

\end{thebibliography}

\begin{appendix}

\section{On the feature tracking method \label{apx:A}}

One of the main concerns of studying small-scale structures with ALMA observations relays on the level of reliability that one should place on the found results. In this particular case, to the sizes of the found features because these are comparable to the ALMA beam during observations. First question that arises is if the features are artefacts produced by either the observation process itself or by the image reconstruction of the interferometric data. To address the former possibility it is safe to assume that the quality assessment of the data by ALMA ensures that the artefacts due to the observation process are minimal. For the latter possibility, we relay on the extensive used of the interferometric techniques over the course of its history, as well as, on the latest studies carried out with solar ALMA observations which have proven ALMA capabilities to distinguish the big picture but also small dynamics structures with an excellent signal to noise ratio. The analysis done in this research has proven the physical nature of the small-scale features found with the algorithm described in Section~\ref{sec:methods}. Therefore, the intention of this appendix is to asses the limitation of the algorithm with regard to the definition of the features borders. To this end, we designed a control experiment consisting of a toy model image filled with Gaussian-like fake features.

The procedure, equivalent for bands 3 and 6, is as follows. Fake 2D Gaussian-like features with known Full Width at Half Maximum (FWHMs) (assumed as the size) were randomly placed in a box with similar size to the respective ALMA band under study. Each 2D Gaussian has a FWHM of 0.5. The true position of each feature within the box and its respective size (FWHM) were stored to be used later on the process. The sizes of the fake features vary randomly between 80\% of the average minor beam size and 5 times the average major beam size. Representing features of sizes in the ranges 1220-8000\,km for band 3 and 400-3100\,km for band 6. Each feature is randomly assigned to a position within the box using a Poisson Disk Sampling method in which the size of the box was defined as 81.6x81.6\,arcsec for band 3 and 33.6x33.6\,arcsec for band 6. It was also set that features should maintain a minimum separation distance between positions of at least the average major beam size. They way of placing all the features together consisted just in summing them up, i.e., when 2 features lie too close their peaks get mixed and they might become indistinguishable. In addition, a random noise with a standard deviation of 5\% the intensity at the FWHM was added to each box. The resulting image is subsequently convolved with the corresponding ALMA beam for each band. The algorithm is then applied to this toy model expecting that it will identify and delimit each feature. 

\begin{figure}[h]
    \centering
    \includegraphics[width=0.48\textwidth]{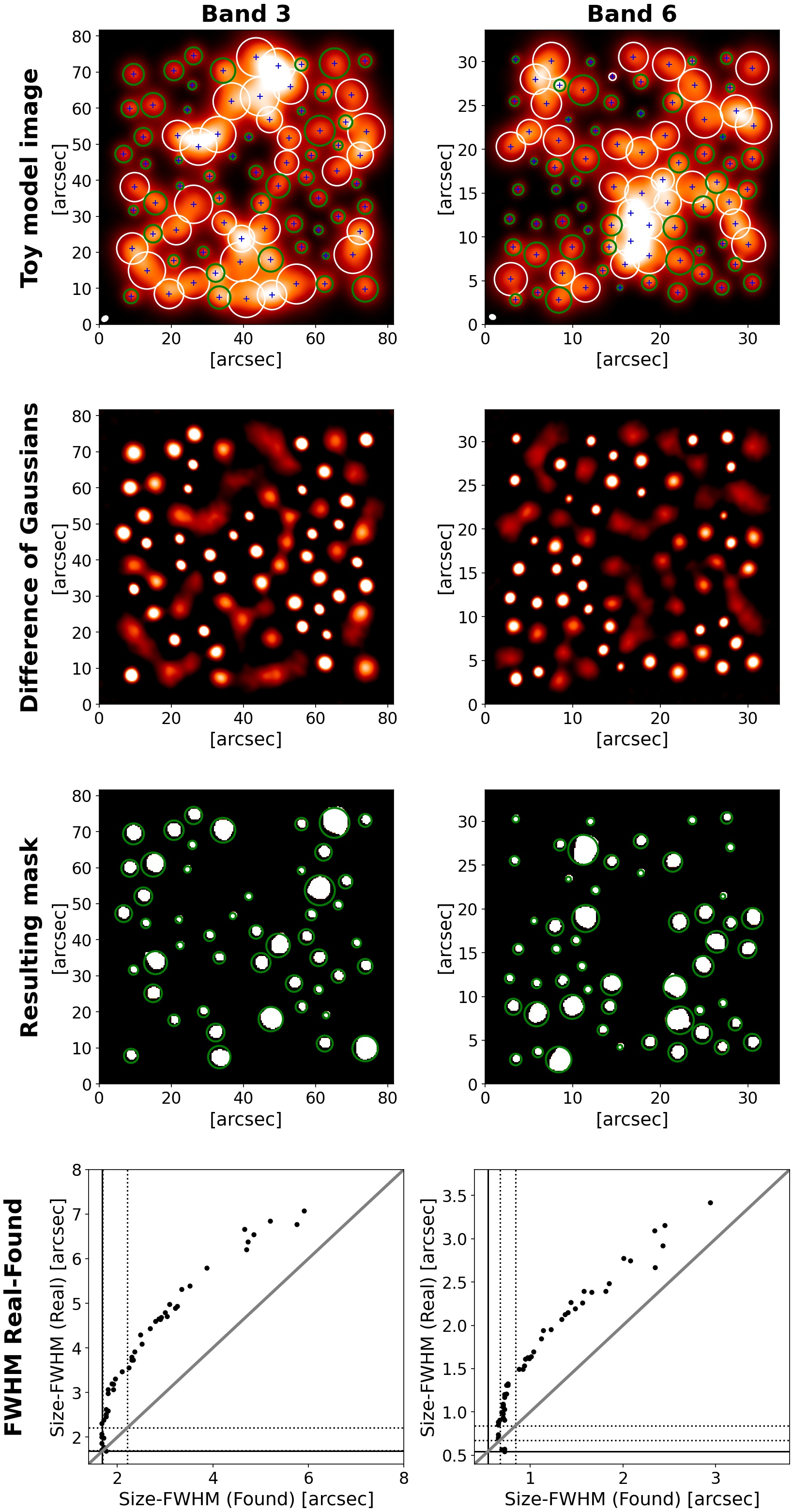}
    \caption{Top panel shows the toy model images for bands 3 and 6. Blue crosses represent the centre location of fake features, green circles shoes those features picked up by the algorithm and white circles those features that were not picked up. The small white ellipses in the bottom left corner show the ALMA beam shape. Second panel shows the difference of Gaussians in each case. Third panel shows the resulting masks and green circles as same as in top panel. Bottom panel shows the relation between real size of the feature and the size found by the algorithm. \label{fig:imgs_fakevert}}
\end{figure}

Figure~\ref{fig:imgs_fakevert} shows in the top panel the toy model images for bands 3 and 6. The white circles represent the FWHM of those features that the algorithm did not find, due to the impossibility of distinguish between two different features whose signals were spatially mixed because of their proximity. The green circles represent those features that were successfully found by the algorithm. The success rate is roughly 60\% for both bands, meaning that the algorithm is capable of picking 6 out of 10 features present in the image. The second panel from the top shows the Difference of Gaussians which is a crucial step in the algorithm as it is used to define the local maxima which would be use to feed the flooding method. The third panel shows the resulting masks, being the green circles same as in top panel. It is possible to perceive that the masks, although having irregular shapes, are still blob-like shapes. The bottom panel shows the relation between the real size of the features, i.e., the FWHM that were kept when the fake features were created, and the sizes found by the algorithm. As a conclusion, it is possible to say that the algorithm is underestimating the size of the features. However, this does not change the results found in this paper, conversely, reinforce the results because staying within the borders of the features ensures that the amount of noise from outside is reduced. Hence, the algorithm has been tested to be an efficient way to delimit small scale structures which in addition to bright features can also be used to find dark features. 

\end{appendix}

\end{document}